\documentclass[journal]{IEEEtran}   

\usepackage{amsmath}
\usepackage{graphicx,psfrag}   
\usepackage{amsmath,amsthm,amsfonts,amssymb,bbm} 
\usepackage[ruled]{algorithm2e}
\usepackage{algpseudocode}
\usepackage{xcolor}
\usepackage{tabularx}          
\usepackage[framed,numbered,autolinebreaks,useliterate]{mcode}
\usepackage[colorlinks,
            linkcolor=black,       
            anchorcolor=black,  
            citecolor=black,        
            ]{hyperref}
\allowdisplaybreaks[3]
\usepackage[latin1]{inputenc}

\theoremstyle{definition}      
\newcommand{\bs}{\mathbf}
\newcommand{\pa}{\mathcal{P}}

\newtheorem{remark}{Remark}

\begin{document}

\title{
A Concise Tutorial on Approximate Message Passing
}

\author{
   Qiuyun Zou and Hongwen Yang
\thanks{
Q. Zou and H. Yang are with Beijing University of Posts and Telecommunications,
Beijing 100876, China (email: qiuyun.zou@bupt.edu.cn; yanghong@bupt.edu.cn).
}
\thanks{
The Matlab code of this paper is available in \url{https://github.com/QiuyunZou/AMPTutorial}.
}
}
\date{\today}
\maketitle

\begin{abstract}
High-dimensional signal recovery of standard linear regression is a key challenge in many engineering fields, such as, communications, compressed sensing, and image processing. The approximate message passing (AMP) algorithm proposed by Donoho \textit{et al} is a computational efficient method to such problems, which can attain Bayes-optimal performance in independent identical distributed (IID) sub-Gaussian random matrices region. A significant feature of AMP is that the dynamical behavior of AMP can be fully predicted by a scalar equation termed station evolution (SE). Although AMP is optimal in IID sub-Gaussian random matrices, AMP may fail to converge when measurement matrix is beyond IID sub-Gaussian. To extend the region of random measurement matrix, an expectation propagation (EP)-related algorithm orthogonal AMP (OAMP) was proposed, which shares the same algorithm with EP, expectation consistent (EC), and vector AMP (VAMP). This paper aims at giving a review for those algorithms. We begin with the worst case, i.e., least absolute shrinkage and selection operator (LASSO) inference problem, and then give the detailed derivation of AMP derived from message passing. Also, in the Bayes-optimal setting, we give the Bayes-optimal AMP which has a slight difference from AMP for LASSO. In addition, we review some AMP-related algorithms: OAMP, VAMP, and Memory AMP (MAMP), which can be applied to more general random matrices.
\end{abstract}


\section{Introduction}
We focus on the sparse signal recovery of the standard linear regression
\begin{align}
\bs{y}=\bs{Hx}+\bs{n},
\end{align}
where $\bs{x}\in \mathbb{R}^N$ is the sparse signal to be estimated, $\bs{H}\in \mathbb{R}^{M\times N} (M\ll N)$ is the measurement matrix which is perfectly known beforehand, $\bs{n}$ is the additive white Gaussian noise with zero mean and covariance $\sigma_w^2$, and $\bs{y}\in \mathbb{R}^M$ is the observation. In the existing works, the sparse signal can be divided into two kinds: one is that $\bs{x}$ is $k$-sparsity but without true distribution, i.e., only $k$ elements of $\bs{x}$ being non-zero, and the other is that $\bs{x}$ is drawn from a specific distribution with sparsity pattern, such as Bernoulli-Gaussian (BG) distribution. Throughout, we focus on the large system limit, in which the dimensions of system tend to infinity $(M,N)\rightarrow \infty$ but the ratio $\alpha=\frac{M}{N}$ is fixed. At the worst case, where prior and likelihood function are both unknown, this sparse inference problem can be formalized as a least absolute shrinkage and selection operator (LASSO) \cite{donoho2006most} inference problem
\begin{align}
\hat{\bs{x}}_{\text{LASSO}}=\underset{\bs{x}}{\arg\min} \ \frac{1}{2}\|\bs{y}-\bs{Hx}\|_2^2+\lambda \|\bs{x}\|_1,
\label{Equ:Lasso}
\end{align}
where $\|\cdot\|_1$ and $\|\cdot\|_2$ are $\ell_1, \ell_2$ norm, respectively, and $\lambda\geq 0$ is the parameter of regularization that balances the sparsity and error of solution. The inference problem above is also known as basis pursuit de-noising (BPDN) inference. Such problem has a mass of  applications in many fields such as compressed sensing \cite{donoho2006most,donoho2006compressed,kabashima2009typical,donoho2010message,donoho2010messageII,donoho2011design}, image processing \cite{beck2009fast,metzler2016denoising}, and sparse channel estimation in wireless communications etc.

\begin{figure*}[!t]
\centering
\includegraphics[width=0.95\textwidth]{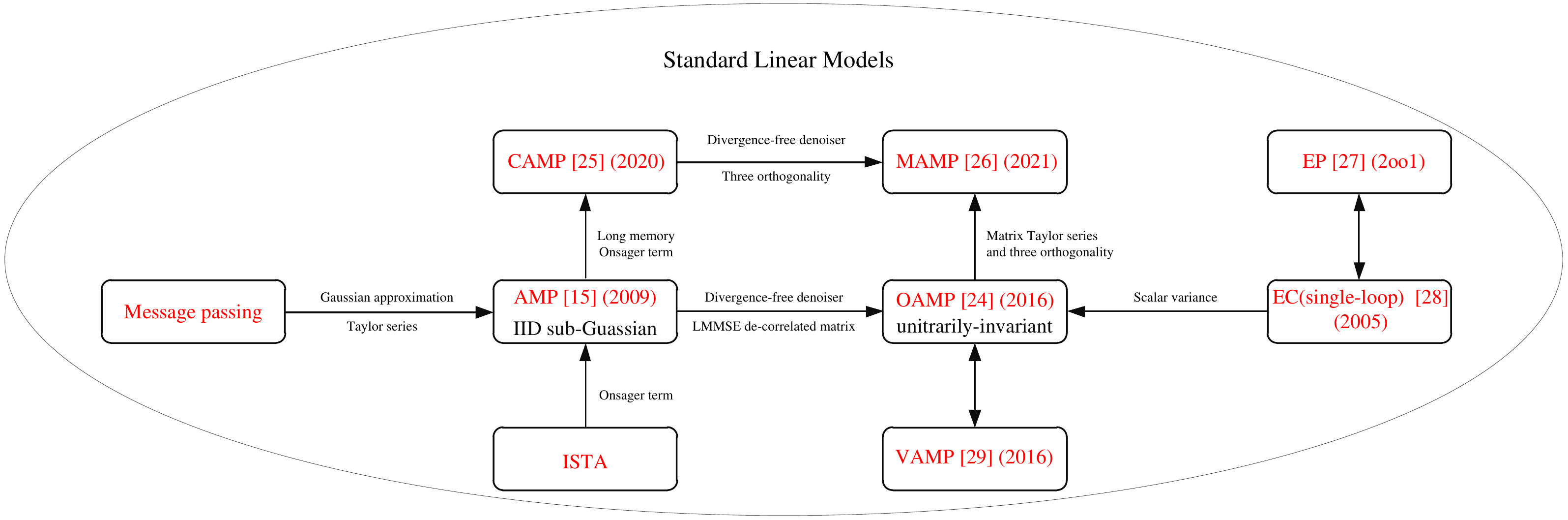}
\caption{The relations between the message passing based algorithms in standard linear regression inference problem.
}
\label{Fig:tree}
\end{figure*}

To solve the LASSO inference problem, there are many kinds of algorithms. For example,
\begin{itemize}
\item
 \textbf{Convex relaxation}. LASSO inference problem is a compound optimization problem involving a smooth function and a non-smooth function such as $\ell_1$ norm regularization. There are a mass of algorithms for compound optimization problem such as sub-gradient method, proximal gradient descent, also known as iterative soft threshold  algorithm (ISTA) \cite{wright2009sparse}, Newton acceleration algorithm, and alternating direction method of multiplies (ADMM) \cite{boyd2011distributed}, etc. Among them, ADMM alternatively optimizes the objective function containing quadric error and the objective function involving $\ell_1$ norm regularization.
\item
 \textbf{Greedy algorithm}. A kind of  alternative method refers to greedy algorithms \cite{blumensath2012greedy} in compressed sensing, such as, match pursuit (MP), orthogonal match pursuit (OMP) \cite{tropp2007signal}, and subspace pursuit (SP) \cite{dai2009subspace}, etc. In those greedy algorithms, they make a `hard' decision based upon some locally optimal optimization criterion. All of those methods can be regarded as a variant of least square. The basic ideal of them is to approximate the signal of interest by selecting the atom or sub-hyperplane from measurement matrix that best matches the residual error of each iteration. Among them, MP projects the residual error of each iteration onto a specific atom, while OMP projects the residual error of each iteration onto a sub-hyperplane from measurement matrix.
\item
\textbf{Bayesian estimation}. The Bayesian estimator \cite[Chapter 10]{kay1993fundamentals} is a kind of algorithm which aims at minimizing the Bayes loss function. According to different Bayes risk functions, the Bayesian estimator can be generally divided into minimum mean square error (MMSE) and maximum a posterior (MAP). In fact, the exact MMSE or MAP is NP-hard problem in general cases. However, there are some algorithms which implement the exact Bayesian estimator iteratively. Among them, the approximate message passing (AMP) \cite{donoho2009message} algorithm, the main focus in this paper, is a celebrated  implementation of Bayes estimation. By postulated posterior/MMSE, in which the postulated prior and likelihood function are different from true ones, AMP can provide the exact sparse solution to LASSO inference problem using Laplace method of integration. In general, we call the algorithm which relies on Bayesian formula as Bayesian algorithms.
 \end{itemize}

On the other hand, in the \textit{Bayes-optimal setting} (may $M\geq N$) where both prior and likelihood function are known,  the MMSE and MAP give a  much better performance than convex relaxation. However, due to high-dimensional integration, the exact MMSE is hard to obtain. Fortunately, some existing works \cite{bayati2011dynamics} showed that AMP can achieve the Bayes-optimal MSE performance but with affordable complexity in independent identical distributed (IID) sub-Gaussian random measurement matrices region \cite{bayati2015universality}. For convenience, we depict Fig.~\ref{Fig:tree} to show the relations between AMP and its related algorithms. The AMP derives from the message passing \cite{kschischang2001factor} algorithm in coding theory, which is also known as belief propagation \cite{kim1983computational} in computer science or cavity method \cite{mezard2009information} in statistic mechanics. The AMP algorithm is closely related to the Thouless-Anderson-Palmer (TAP) \cite{thouless1977solution} equations which is used to approximate marginal moments in large probabilistic models. In \cite{kabashima2003cdma}, the first AMP algorithm was proposed for the code
division multiple access (CDMA) multi-user detection problem.
A significant feature of AMP algorithm is that the dynamic of AMP can be fully predicted by a scalar equation termed state evolution (SE) \cite{bayati2011dynamics}, which is perfectly agree with the fixed point of the exact MMSE estimator using replica method \cite{guo2005randomly}. The AMP algorithm is also related to ISTA, the difference between them is the \textit{Onsager} term, which leads to AMP more faster than ISTA but it doesn't change its fixed points. As the measurement matrix is beyond IID sub-Gaussian region, AMP methods often fail to converge. Beyond IID sub-Gaussian region, the orthogonal AMP (OAMP) \cite{ma2017orthogonal} can be applied to more general unitarily-invariant matrices via the LMMSE de-correlated matrix and divergence-free denoiser, but it should pay more computational complexity due to the matrix inversion. To balance the complexity and region of random measurement matrix, recently, some long memory algorithms such as convolutional AMP (CAMP) \cite{takeuchi2021bayes}, and memory AMP \cite{liu2021memory} were proposed.
Different from OAMP, CAMP only modifies the Onsager term of AMP. The Onsager term of CAMP includes all proceeding messages to ensure the Gaussianity of input signal of denoiser. However, CAMP may fail to converge in the case of large condition number. Following CAMP and OAMP, the MAMP algorithm applies finite terms of matrix Taylor series to approximate matrix inversion of OAMP and involves all previous messages to ensure three orthogonality.

Another efficient algorithm related to AMP is called expectation propagation (EP) \cite{minka2001family}. EP is earlier than AMP, which approximates the factorable factors by choosing a distribution from Gaussian family via minimizing Kullback-Leibler (KL) divergence. Some EP-related methods refer to expectation consistent (EC) \cite[Appendix D]{opper2005expectation} (single-loop), OAMP \cite{ma2017orthogonal}, and vector AMP (VAMP) \cite{rangan2019vector}. They were proposed independently in different manners but share the same algorithm. Actually, EP/EC (single-loop) have a slight difference from OAMP/VAMP, since EP/EC has the element-wise variances and they can be reduced to OAMP/VAMP by taking the mean operation for element-wise variance.  Among them, EC approximation is based on the minimum Gbiss free energy. It means that those methods can be regarded as an example of solving the fixed point of Gbiss free energy. Almost at the same time as OAMP, the VAMP was proposed using a EP-type message passing and the dynamic of VAMP was rigorously analyzed in \cite{rangan2019vector}. Recently, \cite{zhang2020identical} proved that VAMP and AMP have identical fixed points in their state
evolutions in their overlapping random matrices. We also note that under the mismatch case \cite{takahashi2020macroscopic}, where the prior and likelihood function applied to the inference problem are different from the true prior and likelihood function, the AMP as well as its related algorithms  may not converge although the corresponding SE converges to a fixed point predicted by replica method. Actually, AMP for LASSO is one case of mismatched model, but its convergence is guaranteed due to convex nature of LASSO \cite{gerbelot2020asymptotic}. The failure of AMP can occur when the mismatched models are defined by \textit{non-convex} cost function \cite{obuchi2019cross}.

Besides, there are some algorithms that extend AMP to more general models beyond standard liner model. In \cite{rangan2011generalized}, a generalized AMP (GAMP) algorithm was proposed for generalized linear model which allows an arbitrary row-wise mapping. A concise derivation of GAMP using EP projection can be found in \cite{meng2015expectation,zou2018concise}. Further, Park \textit{et al} \cite{parker2014bilinear} developed bilinear GAMP (BiG-AMP) which extends the GAMP algorithm to bilinear model in which both the signal of interest and measurement matrix are unknown. Recent works showed that the BiG-AMP can be obtained by Plefka-Georges-Yedidia method \cite{maillard2019high,maillard2021perturbative}.
Following VAMP,  \cite{schniter2016vector,he2017generalized} developed a generalized linear model VAMP (GLM-VAMP) algorithm by constructing an equivalent linear model. Compared to GAMP, GLM-VAMP can be applied to more general random matrices but needs to pay more computational complexity. Similar to GLM-VAMP, a generalized version of MAMP was proposed in \cite{tian2021generalized}.  Beyond single-layer model, some extensions of AMP in multi-layer regions can be found in \cite{manoel2017multi,fletcher2018inference,pandit2019asymptotics,zou2021multi}. However, those algorithms are out of the scope of this paper.

Although AMP and its related methods have attracted a lot of attention in many engineering fields, there still isn't a tutorial that gives a clear line to summarize them and provides concise derivations. That is the purpose of this paper. For that purpose, we begin with the LASSO inference problem, which is original goal of AMP. By Laplace method of integration, the LASSO inference problem can be converted into the limit of postulated MMSE estimator. Using factor graph representation and message passing, we give the detailed derivation of AMP for LASSO. And then we move to the Bayes-optimal setting, which is more attractive and common in some engineering fields, such as wireless communications. Beyond IID sub-Gaussian random matrices, we review several extensions of AMP: OAMP, VAMP, and MAMP, and illustrate their relations and differences.

\textit{Notations}: Throughout, we use $\bs{x}$ and $\bs{X}$ to denote column vector and matrix, respectively. $(\cdot)^{\text{T}}$ denotes transpose operator such as $\bs{X}^{\text{T}}$. $\text{Tr}(\bs{A})$ denotes the trace of square matrix $\bs{A}$. $\overset{\text{a.s.}}{=}$ means equal almost sure. Given the original signal $\bs{x}$ and its estimator $\hat{\bs{x}}$, the normalized MSE (NMSE) is defined as $\text{NMSE}(\bs{x})=\frac{\|\hat{\bs{x}}-\bs{x}\|_2^2}{\|\bs{x}\|_2^2}$ with $\|\cdot\|_2$ being $\ell_2$ norm. We apply $\mathcal{N}(x|a,A)$ to denote a Gaussian probability density function with mean $a$ and variance $A$ described by:
$$\mathcal{N}(x|a,A)=\frac{1}{\sqrt{2\pi A}}\exp \left[-\frac{(x-a)^2}{2A}\right].$$
$\mathcal{BG}(\mu, \rho)$ is a Bernoulli Gaussian distribution: $\mathcal{BG}(\mu, \rho)=\rho\mathcal{N}(x|\mu, \rho^{-1})+(1-\rho)\delta(x)$.

\section{Approximate Message Passing}
\subsection{Iterative Soft Threshold Algorithm}
Before introducing AMP algorithm, we first review a AMP related algorithm: ISTA. Recalling that the term $f(\bs{x})=\frac{1}{2}\|\bs{y}-\bs{Hx}\|_2^2$ in (\ref{Equ:Lasso}) is continuous and derivative while the second term $g(\bs{x})=\lambda\|\bs{x}\|_1$ is not differentiable at $\bs{x}=\bs{0}$. The minimization of $f(\bs{x})$ can be achieved by gradient descent
\begin{align}
\!\!\!\!\!\hat{\bs{x}}^{(t)}=\underset{\bs{x}}{\arg \min}\ \frac{1}{2\alpha_t}\|\bs{x}-(\hat{\bs{x}}^{(t-1)}-\alpha_{t-1}\nabla f(\hat{\bs{x}}^{(t-1)}))\|_2^2,
\label{Equ:A1}
\end{align}
where $\alpha_t$ is step size  and $\hat{\bs{x}}^{(t)}$ is the estimator of $\bs{x}$ at $t$-iteration. Adding $\ell_1$ norm regularization, (\ref{Equ:A1}) becomes
\begin{align}
\hat{\bs{x}}^{(t)}=\underset{\bs{x}}{\arg \min} \frac{1}{2\alpha_t}\|\bs{x}-(\hat{\bs{x}}^{(t-1)}-\alpha_{t-1}\nabla f(\hat{\bs{x}}^{(t-1)}))\|_2^2+\lambda \|\bs{x}\|_1.
\end{align}

\begin{figure}[!t]
\centering
\includegraphics[width=0.45\textwidth]{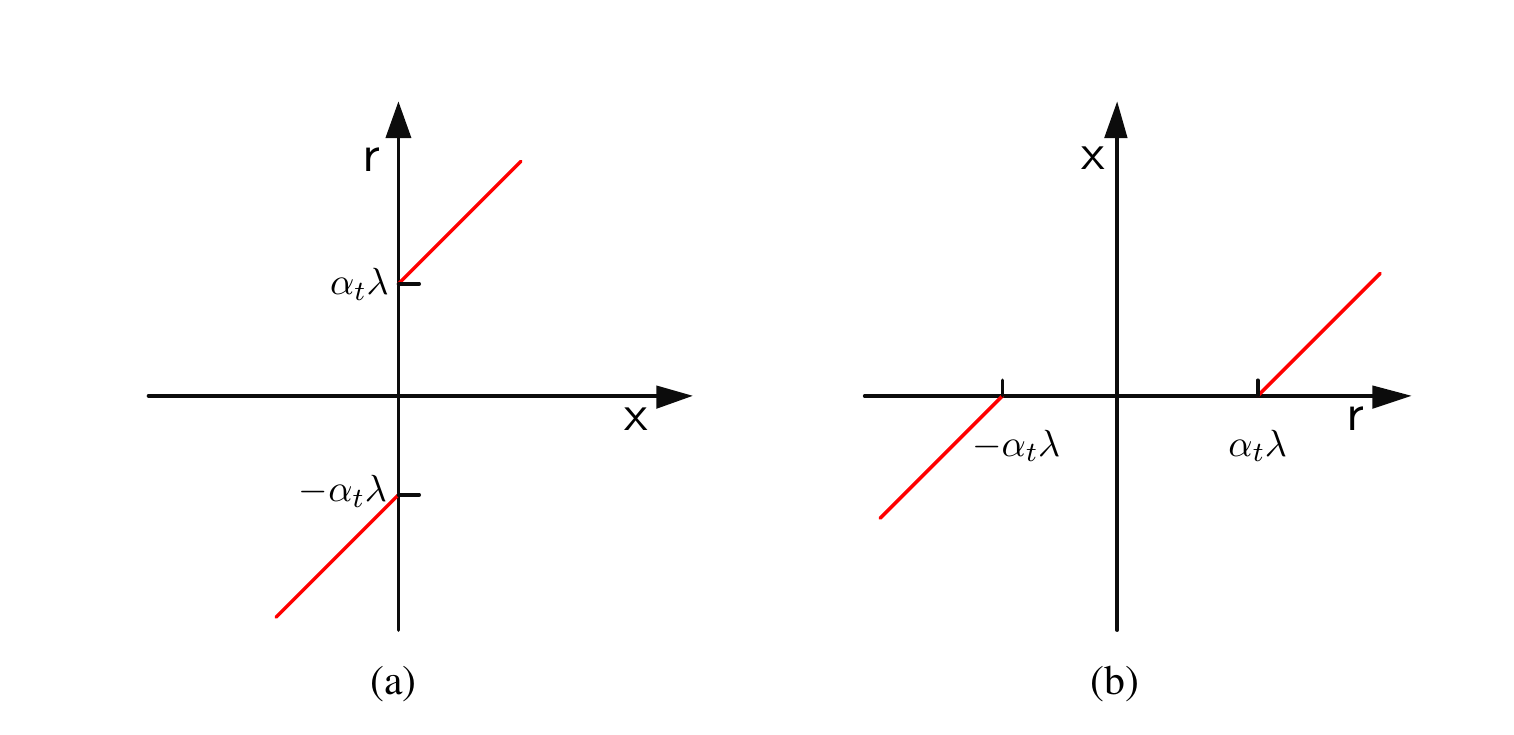}
\caption{(a) $\textsf{x}$-$\textsf{z}$ coordinate axis; (b) $\textsf{z}$-$\textsf{x}$ coordinate axis.
}
\label{Fig:axes}
\end{figure}

Defining $\bs{r}^{(t)}=\hat{\bs{x}}^{(t)}-\alpha_{t}\nabla f(\hat{\bs{x}}^{(t)})=\hat{\bs{x}}^{(t)}+\alpha_t\bs{H}^{\text{T}}(\bs{y}-\bs{H}\hat{\bs{x}}^{(t)})$, the equation above becomes
\begin{align}
\forall i:\quad  \hat{x}_i^{(t)}=\underset{x_i}{\arg \min} \left\{\frac{1}{2\alpha_t}(x_i-r_i^{(t-1)})^2+\lambda|x_i|\right\}.
\label{Equ:B1}
\end{align}
Zeroing the gradients w.r.t. $x_i$ yields
$
r_i^{(t-1)}=\hat{x}_i^{(t)}+\alpha_t\lambda \text{sign}(\hat{x}_i^{(t)})
$
. Then swapping the axes (see Fig~\ref{Fig:axes}) gets
\begin{align}
\hat{x}_i^{(t)}=\text{sign}(r_i^{(t-1)})\max(|r_i^{(t-1)}|-\alpha_t\lambda,0).
\label{Equ:B2}
\end{align}
Totally, the ISTA  is summarized as
\begin{subequations}
\begin{align}
\bs{r}^{(t)}&=\hat{\bs{x}}^{(t)}+\alpha_t \bs{H}^{\text{T}}(\bs{y}-\bs{H}\hat{\bs{x}}^{(t)}),\\
\hat{\bs{x}}^{(t+1)}&=\text{sign}(\bs{r}^{(t)})\max(|\bs{r}^{(t)}|-\alpha_t\lambda, 0).
\end{align}
\label{ISTA}
\end{subequations}

To in line with AMP, let's define $\bs{z}^{(t)}=\bs{y}-\bs{H}\hat{\bs{x}}^{(t)}$ and $\eta(\bs{r}^{(t)},\alpha_t\lambda )=\text{sign}(\bs{r}^{(t)})\max(|\bs{r}^{(t)}|, \alpha_t\lambda)$. The ISTA algorithm can be written as
\begin{subequations}
\begin{align}
\bs{z}^{(t)}&=\bs{y}-\bs{H}\hat{\bs{x}}^{(t)},\\
\hat{\bs{x}}^{(t+1)}&=\eta(\hat{\bs{x}}^{(t)}+\bs{H}^{\text{T}}\bs{z}^{(t)}, \lambda),
\end{align}
\label{Alg:ISTA}
\end{subequations}
\!\!where the step size $\alpha_t$ is set to $\alpha_t=1$. However, in practical,  $\alpha_t$ may cause the algorithm to diverge and actually $\alpha_t\in [0.1, 0.35]$ is appropriate in our simulation. The complexity of ISTA is dominated by the matrix multiplication with the cost of $\mathcal{O}(MN)$.  However, the convergence speed of ISTA is too slow. To improve the convergence speed of ISTA, the fast ISTA (FISTA) \cite{beck2009fast} was proposed. The FISTA is beyond the scope of this paper. We only post it as below
\begin{subequations}
\begin{align}
\hat{\bs{x}}^{(t)}&=\hat{\bs{x}}^{(t-1)}+\frac{t-2}{t+1}(\hat{\bs{x}}^{(t-1)}-\hat{\bs{x}}^{(t-2)}),\\
\bs{z}^{(t)}&=\bs{y}-\bs{H}\hat{\bs{x}}^{(t)},\\
\hat{\bs{x}}^{(t+1)}&=\eta(\hat{\bs{x}}^{(t)}+\bs{H}^{\text{T}}\bs{z}^{(t)}).
\end{align}
\label{FISTA}
\end{subequations}
\!\!\!Comparing FISTA in (\ref{FISTA}) with ISTA in (\ref{ISTA}), the difference between them is that the term $\hat{\bs{x}}^{(t)}$ is constructed from two previous results.

\subsection{AMP for LASSO}
The AMP algorithm \cite{donoho2009message} posted below is related to ISTA
\begin{subequations}
\begin{align}
\nonumber
\bs{z}^{(t)}&=\bs{y}-\bs{H}\hat{\bs{x}}^{(t)}\\
&\quad +\frac{1}{\alpha}\bs{z}^{(t-1)}\left<\eta'_{t-1}(\hat{\bs{x}}^{(t-1)}+\bs{H}^{\text{T}}\bs{z}^{(t-1)})\right>,\\
\hat{\bs{x}}^{(t+1)}&=\eta_t(\hat{\bs{x}}^{(t)}+\bs{H}^{\text{T}}\bs{z}^{(t)}),
\end{align}
\label{Alg:AMP}
\end{subequations}
\!\!where $\left<\cdot \right>$ is empirical mean such as $\left<\bs{x}\right>=\frac{1}{N}\sum_{i=1}^N x_i$ and $\eta_{t-1}'(\bs{r})$ is the partial derivation of $\eta_{t-1}(\cdot)$ w.r.t. $\bs{r}$.

\begin{figure}[!t]
\centering
\includegraphics[width=0.43\textwidth]{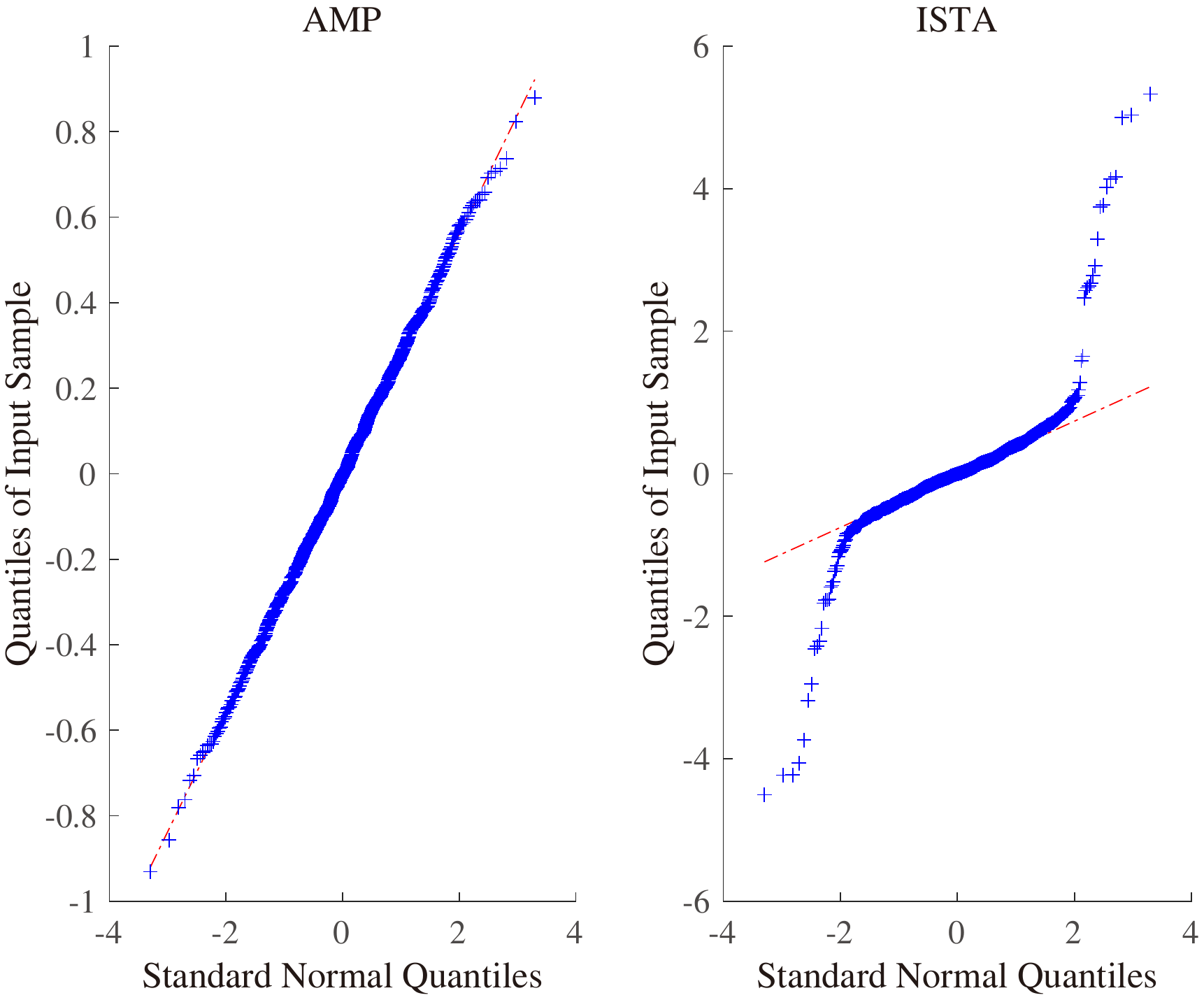}
\caption{QQplot comparing the distribution of input error $\bs{r}^{(t)}-\bs{x}$ of AMP and ISTA at $t=5$. The system setups are similar to that of Fig.~\ref{EXP1}. The blue points match the red line better, the closer the input error is to the Gaussian distribution. Notice that the input error of AMP remains Gaussianity due to Onsager term.
}
\label{EXP1_2}
\vspace{+0.4cm}
\centering
\includegraphics[width=0.43\textwidth]{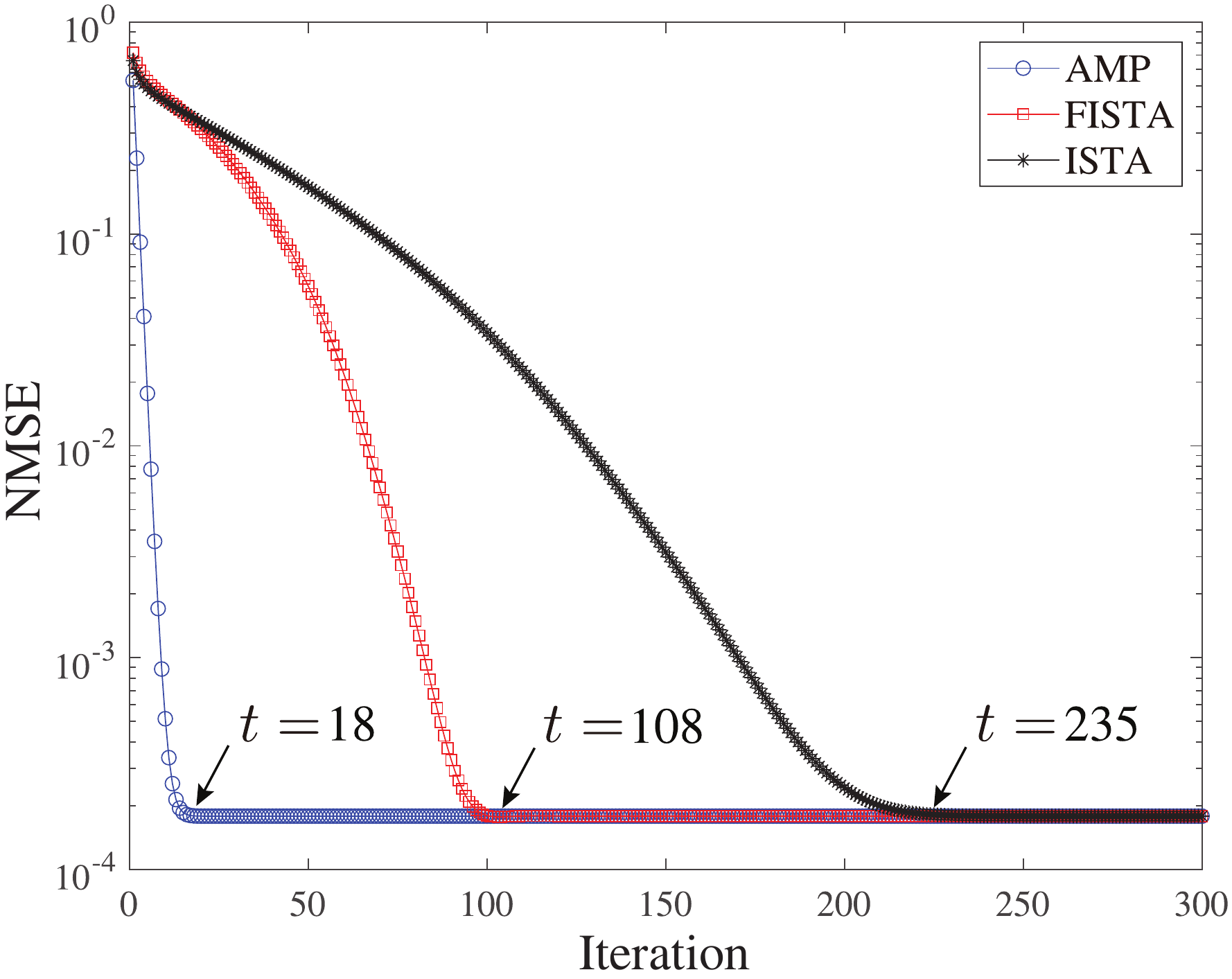}
\caption{Comparison of AMP, FISTA, and ISTA for LASSO inference problem. $\bs{H}$ has IID Gaussian entries with zero mean and $1/M$ variance. $N=1024$, $M=512$, $\alpha=\frac{M}{N}=\frac{1}{2}$, and $\lambda=0.05$. $\text{SNR}=1/\sigma_w^2=50\text{dB}$. $\bs{x}$ has IID BG entries following $\mathcal{BG}(0,0.05)$. The step size $\alpha_t=0.35$ and $\alpha_t=0.2$ are applied to ISTA and FISTA, respectively.
}
\label{EXP1}
\end{figure}

Compared to ISTA algorithm in (\ref{Alg:ISTA}), the key difference between AMP and ISTA is the \textit{Onsager} term $\frac{1}{\alpha}\bs{z}^{(t-1)}\left<\eta_{t-1}'(\bs{x}^{(t-1)}+\bs{H}^{\text{T}}\bs{z}^{(t-1)})\right>$. This term can improve the convergence speed of ISTA but does not change its fixed point. Essentially, this term ensures that the input $\bs{r}^{(t)}$ of denoiser $\eta(\cdot)$ can be expressed as  the original signal adding an additive Gaussian noise (Gaussianity, see Fig.~\ref{EXP1_2}) and it leads to faster convergence than ISTA. As shown in Fig.~\ref{EXP1}, we compare per-iteration NMSE behavior of the AMP with ISTA and FISTA. From Fig.~\ref{EXP1}, we can see that AMP converges with $t=18$ iterations which is far small than  FISTA ($t=108$) and  ISTA ($t=235$). Be aware, in ISTA, one should adjust the step size $\alpha_t$ to ensure the convergence but the step size is unnecessary to AMP. In addition, an appropriate step size ensures the algorithm to converge but does not change the fixed point. The below is the detailed derivation to obtain AMP for LASSO inference problem.

As shown in \cite[Appendix D]{rangan2012asymptotic}, \cite{merhav2010statistical}, the LASSO inference problem can be expressed as the limit of the postulated MMSE estimator using Laplace method of integration
\begin{align}
\nonumber
\hat{\bs{x}}
&=\lim_{\beta\rightarrow \infty}\int \bs{x}\underbrace{\frac{1}{Z_{\beta}^{\text{pos}}}\exp \left[-\beta\left(\frac{1}{2}\|\bs{y}-\bs{Hx}\|_2^2+\lambda \|\bs{x}\|_1\right)\right]}_{q(\bs{x}|\bs{y})}\text{d}\bs{x}
\\
&=\underset{\bs{x}}{\arg\min} \left\{\frac{1}{2}\|\bs{y}-\bs{Hx}\|_2^2+\lambda \|\bs{x}\|_1\right\},
\label{Chapter_AMP:pos}
\end{align}
where $Z^{\text{pos}}_{\beta}$ is the normalization constant. Using Bayes' rules, the postulated posterior $q(\bs{x}|\bs{y})$ in (\ref{Chapter_AMP:pos}) is expressed as
\begin{align*}
q(\bs{x}|\bs{y})&=\frac{1}{q(\bs{y})}{q(\bs{x})q(\bs{y}|\bs{x})},\\
q(\bs{x})&=\frac{1}{Z^{\text{pri}}_{\beta}}\exp (-\beta\lambda \|\bs{x}\|_1),\\
q(\bs{y}|\bs{x})&=\frac{1}{Z^{\text{lik}}_{\beta}}\exp \left(-\frac{\beta}{2}\|\bs{y}-\bs{Hx}\|_2^2\right),
\end{align*}
where $q(\bs{y})$, $Z^{\text{pri}}_{\beta}$, and $Z^{\text{lik}}_{\beta}$ are normalization constants, and $q(\bs{x})$ is the postulated prior while $q(\bs{y}|\bs{x})$ is the postulated likelihood function. The postulated likelihood function can also be formalized as $q(\bs{y}|\bs{x})=\prod_{a=1}^M \mathcal{N}(y_a|\sum_{i=1}^Nh_{ai}x_i, \frac{1}{\beta})$.

\begin{figure}[!t]
\centering
\includegraphics[width=0.25\textwidth]{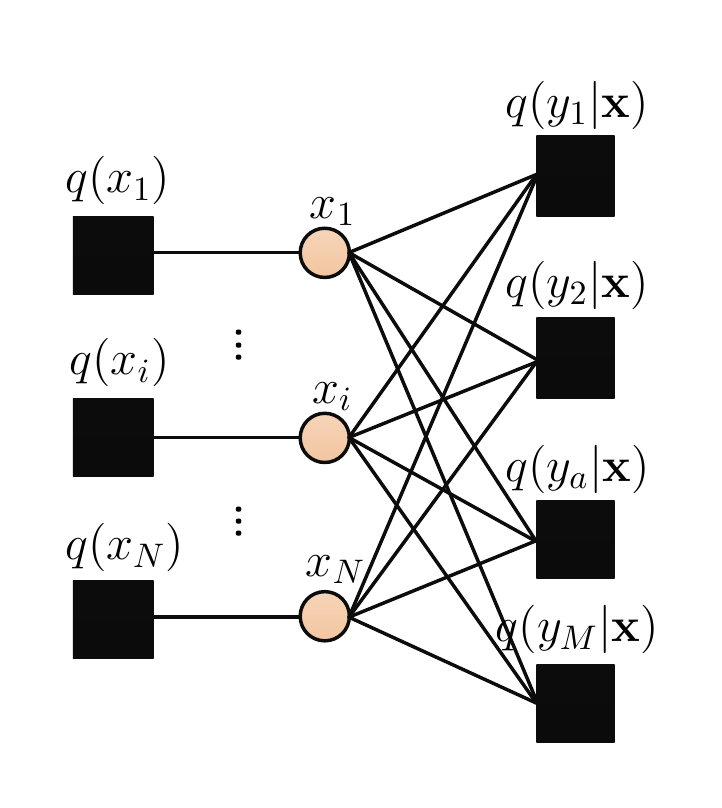}
\caption{Factor graph of postulated posterior $q(\bs{x}|\bs{y})$ defined in (\ref{Chapter_AMP:pos}), where $q(x_i)\propto e^{-\beta|x_i|}$ and $q(y_a|\bs{x})=\mathcal{N}(y_a|\sum_{i=1}^N h_{ai}x_i, 1/\beta)$. The square denotes the factor node (e.g. $q(x_i)$) while the circle denotes the variable node (e.g. $x_i$). The messages delivers between factor nodes and variable nodes via their edges.
}
\label{Fig:SLM}
\end{figure}

The factor graph of postulated posterior defined in (\ref{Chapter_AMP:pos}) is depicted in Fig.~\ref{Fig:SLM}. For basis of factor graph and message passing, we suggest \cite[Chapter 2]{richardson2008modern} for more details. From this figure, the messages are addressed as
\begin{subequations}
\begin{align}
\mu_{i\rightarrow a}^{(t+1)}(x_i)&\propto e^{-\beta\lambda |x_i|}\prod_{b\ne a}^M \mu_{i\leftarrow b}^{(t)}(x_i),\\
\mu_{i\leftarrow a}^{(t)}(x_i)&\propto \int q(y_a|\bs{x})\prod_{j\ne i}^N\mu_{j\rightarrow a}^{(t)}(x_j)\text{d}\bs{x}_{\backslash i},
\end{align}
\label{Chapter_AMP:MP}
\end{subequations}
\!\!where $\bs{x}_{\backslash i}$ is $\bs{x}$ expect $x_i$, $\mu_{i\rightarrow a}^{(t+1)}(x_i)$ is the message from variable node $x_i$ to factor node $q(y_a|\bs{x})$, $\mu_{i\leftarrow a}^{(t)}(x_i)$ is the message in opposite direction at $t$-iteration, and superscript $t$ denotes the number of iteration. It is worth noting that at $t$-iteration, the marginal posterior $\pa(x_i|\bs{y})$ can be approximated by
\begin{align}
\hat{q}^{(t+1)}(x_i|\bs{y})=\frac{e^{-\beta\lambda |x_i|}\prod_{a=1}^M \mu_{i\leftarrow a}^{(t)}(x_i)}{\int e^{-\beta\lambda |x_i|}\prod_{a=1}^M \mu_{i\leftarrow a}^{(t)}(x_i) \text{d}x_i},
\label{Equ:posterior}
\end{align}
while the mean of the approximated posterior $\hat{q}^{(t+1)}(x_i|\bs{y})$ will serve as an approximation of MMSE estimator.

To reduce the complexity of sum-product message passing shown in (\ref{Chapter_AMP:MP}), we first simplify the message $\mu_{i\leftarrow a}^{(t)}(x_i)$ as below
\begin{align}
\nonumber
&\mu_{i\leftarrow a}^{(t)}(x_i)\\
\nonumber
&\propto \int_{\bs{x}_{\backslash i}} \int_{z_a}q(y_a|z_a)\delta(z_a-\sum_{k=1}^Nh_{ak}x_k)\text{d}z_a\prod_{j\ne i}^N\mu_{j\rightarrow a}^{(t)}(x_j)\text{d}\bs{x}_{\backslash i}\\
&\propto \int_{z_a}q(y_a|z_a)\mathbb{E} \left\{\delta \left(z_a-\sum_{j\ne i}h_{aj}x_j-h_{ai}x_i\right)\right\}\text{d}z_a
\label{Chapter_AMP:A1},
\end{align}
where the expectation is over $\prod_{j\ne i}^N\mu_{j\rightarrow a}^{(t)}(x_j)$.
We define random variable (RV) $\zeta_{i\leftarrow a}^{(t)}$ associated with $z_a$ and $\xi_{j\rightarrow a}^{(t)}$ following $\mu_{j\rightarrow a}^{(t)}(x_j)$ associated with $x_j$. Denote the mean and variance of $\xi_{j\rightarrow a}^{(t)}$ as $\hat{x}_{j\rightarrow a}^{(t)}$ and $\hat{v}_{j\rightarrow a}^{(t)}/\beta$, respectively. From (\ref{Chapter_AMP:A1}), as the dimension $N$ tends to infinity, using central limit (CLT) theorem the RV $\zeta_{i\leftarrow a}^{(t)}$ converges to a Gaussian RV with mean and variance
\begin{align}
\mathbb{E}\{\zeta_{i\leftarrow a}^{(t)}\}=Z_{i\leftarrow a}^{(t)}+h_{ai}x_i, \ \text{Var}\{\zeta_{i\leftarrow a}^{(t)}\}=\frac{1}{\beta}V_{i\leftarrow a}^{(t)},
\end{align}
where
\begin{align}
Z_{i\leftarrow a}^t&=\sum_{j\ne i}h_{aj}\hat{x}_{j\rightarrow a}^t, \quad V_{i\leftarrow a}^t=\sum_{j\ne i}|h_{aj}|^2\hat{v}_{j\rightarrow a}^t.
\end{align}
Based on this Gaussian approximation, the term $\mathbb{E} \{\delta (z_a-\sum_{j\ne i}h_{aj}x_j-h_{ai}x_i)\}$ in (\ref{Chapter_AMP:A1}) is replaced by $\mathcal{N}(z_a|h_{ai}x_i+Z_{i\leftarrow a}^{(t)}, \frac{1}{\beta}V_{i\leftarrow a}^{(t)})$. By Gaussian reproduction lemma\footnote{
$\mathcal{N}(x|a,A)\mathcal{N}(x|b, B)=\mathcal{N}(x|c,C)\mathcal{N}(0|a-b, A+B)$ with $C=(A^{-1}+B^{-1})^{-1}$ and $c=C\left(\frac{a}{A}+\frac{b}{B}\right)$
}, the message $\mu_{i\leftarrow a}^{(t)}(x_i)$ is approximated as
\begin{align}
\nonumber
\mu_{i\leftarrow a}^{(t)}(x_i)
&\propto \mathcal{N}\left(0|y_a-h_{ai}x_i-Z_{i\leftarrow a}^{(t)}, \frac{1}{\beta}(1+V_{i\leftarrow a}^{(t)})\right)\\
&\propto \mathcal{N}\left(x_i|\frac{y_a-Z_{i\leftarrow a}^t}{h_{ai}},\frac{1+V_{i\leftarrow a}^{(t)}}{\beta|h_{ai}|^2}\right).
\label{Equ:E1}
\end{align}
In the sequel, the mean and variance of $\mu_{i\leftarrow a}^{(t)}(x_i)$ are defined and evaluated as
\begin{align}
\hat{x}_{i\leftarrow a}^{(t)}=
\frac{y_a-Z_{i\leftarrow a}^{(t)}}{h_{ai}}, \quad  \hat{v}_{i\leftarrow a}^{(t)}=\frac{1+V_{i\leftarrow a}^{(t)}}{\beta|h_{ai}|^2}.
\end{align}
Be aware the equation (\ref{Equ:E1}) is mathematically invalid as $h_{ai}=0$. However, in the rest of this section we will show that several zero elements in $\bs{H}$ has no effect on the final result.

Let's move to calculate the message $\mu_{i\rightarrow a}^{(t+1)}(x_i)$ in (\ref{Chapter_AMP:MP}) based on the approximated result above. Applying Gaussian reproduction property, the term $\prod_{b\ne a}^M \mu_{i\leftarrow b}^{(t)}(x_i)$ in $\mu_{i\rightarrow a}^{(t+1)}(x_i)$ is proportion to
\begin{align}
\prod_{b\ne a}^M \mu_{i\leftarrow b}^{(t)}(x_i)\propto \mathcal{N}(x_i|r_{i\rightarrow a}^{(t)}, \Sigma_{i\rightarrow a}^{(t)}),
\end{align}
where
\begin{align}
\Sigma_{i\rightarrow a}^{(t)}&=\left(\sum_{b\ne a}\frac{|h_{bi}|^2}{1+V_{i\leftarrow b}^{(t)}}\right)^{-1},\\
r_{i\rightarrow a}^{(t)}&=\Sigma_{i\rightarrow a}^t\sum_{b\ne a}\frac{h_{bi}^{*}(y_b-Z_{i\leftarrow b}^{(t)})}{1+V_{i\leftarrow b}^{(t)}}.
\end{align}
Note that several zero value elements in $\bs{H}$ have no effect on $\Sigma_{i\rightarrow a}^{(t)}$, $r_{i\rightarrow a}^{(t)}$ as well as rest parameters in the derivation of AMP.

As a result, the message $\mu_{i\rightarrow a}^{(t+1)}(x_i)$ is approximated as the product of a Laplace prior and a Gaussian likelihood function
\begin{align}
\mu_{i\rightarrow a}^{(t+1)}(x_i)=\frac{1}{Z_{\beta}}e^{-\beta\lambda |x_i|}\mathcal{N}(x_i|r_{i\rightarrow  a}^{(t)}, \Sigma_{i\rightarrow a}^{(t)}),
\end{align}
where $Z_{\beta}$ is normalized constant.

For convenience, define a distribution
\begin{align}
f_{\beta}(x; r, \Sigma)=\frac{1}{Z_{\beta}}\exp \left[-\beta\left(\lambda |x|+\frac{1}{2\Sigma}(x-r)^2\right)\right],
\end{align}
and its mean and variance
\begin{align}
\textsf{F}_{\beta}(x;r,\Sigma)&=\int x f_{\beta}(x; r, \Sigma)\text{d}x,
\\
\textsf{G}_{\beta}(x; r, \Sigma)&=\int x^2 f_{\beta}(x;r, \Sigma)\text{d}x-|\textsf{F}_{\beta}(x;r,\Sigma)|^2.
\end{align}
The mean and variance of the message $\mu_{i\rightarrow a}^{(t+1)}(x_i)$ are represented as
\begin{align}
\hat{x}_{i\rightarrow a}^{(t+1)}&=\textsf{F}_{\beta}(x_i;r_{i\rightarrow a}^{(t)},\Sigma_{i\rightarrow a}^{(t)}),
\label{Equ:F1}\\
\hat{v}_{i\rightarrow a}^{(t+1)}&=\beta \textsf{G}_{\beta}(x_i;r_{i\rightarrow a}^{(t)},\Sigma_{i\rightarrow a}^{(t)}).
\label{Equ:Fa}
\end{align}
Recalling the approximated posterior $\hat{q}^{(t+1)}(x_i|\bs{y})$ in (\ref{Equ:posterior}), we define
\begin{align}
\Sigma_i^{(t)}&=\left(\sum_{a=1}^M\frac{|h_{ai}|^2}{1+V_{i\leftarrow a}^{(t)}}\right)^{-1},\\
r_i^{(t)}&=\Sigma_{i}^{(t)}\sum_{a=1}^M\frac{h_{ai}^{*}(y_a-Z_{i\leftarrow a}^{(t)})}{1+V_{i\leftarrow a}^{(t)}}
\label{Equ:F7}.
\end{align}
The term $\prod_{a=1}^M \mu_{i\leftarrow a}^{(t)}(x_i)$ is proportion to $\mathcal{N}(x_i|r_i^{(t)}, \Sigma_i^{(t)})$. Accordingly, the mean and variance of approximated posterior $\hat{q}^{(t+1)}(x_i|\bs{y})$ can be denoted as
\begin{align}
\hat{x}_{i}^{(t+1)}&=\textsf{F}_{\beta}(x_i;r_{i}^{(t)},\Sigma_{i}^{(t)}),\\
\hat{v}_{i}^{(t+1)}&=\beta \textsf{G}_{\beta}(x_i;r_{i}^{(t)},\Sigma_{i}^{(t)}).
\end{align}

Also define
\begin{align}
Z_a^{(t)}&=\sum_{i=1}^Nh_{ai}\hat{x}_{i\rightarrow a}^{(t)}
\label{Equ:Za}\\
V_{a}^{(t)}&=\sum_{i=1}|h_{ai}|^2\hat{v}_{i\rightarrow a}^t\approx V_{i\leftarrow a}^{(t)}
\label{Equ:Va}
\end{align}
where $V_a^{(t)}=V_{i\leftarrow a}^{(t)}$ holds by ignoring infinitesimal.

Applying first-order Taylor series\footnote
{
$f(x+\triangle x, y+\triangle y)=f(x,y)+\triangle xf'_x(x,y)+\triangle y f_y'(x,y)$, where $f_x'$ and $f_y'$ are the partial derivation of $f(x,y)$ w.r.t. $x$ and $y$, respectively.
} to $\hat{x}_{i\rightarrow a}^{(t+1)}$ in (\ref{Equ:F1}), we have
\begin{align}
\nonumber
\hat{x}_{i\rightarrow a}^{(t+1)}
&\approx \hat{x}_i^{(t+1)}+\triangle r \frac{\partial }{\partial r}\textsf{F}_{\beta}(x_i;r_i^{(t)},\Sigma_i^{(t)})\\
&\quad +\triangle {\Sigma} \frac{\partial }{\partial \Sigma}\textsf{F}_{\beta}(x_i;r_i^{(t)},\Sigma_i^{(t)}),
\label{Equ:F2}
\end{align}
where
\begin{align}
\nonumber
\triangle \Sigma
&=\Sigma_{i\rightarrow a}^{(t)}-\Sigma_i^{(t)}\\
\nonumber
&=\frac{\frac{|h_{ai}|^2}{1+V_a^{(t)}}}{\left(\sum_{a=1}^M\frac{|h_{ai}|^2}{1+V_{i\leftarrow a}^{(t)}}\right)\left(\sum_{b\ne a}^M\frac{|h_{bi}|^2}{1+V_{i\leftarrow b}^{(t)}}\right)}\\
&\approx 0,\\
\nonumber
\triangle r
&=r_{i\rightarrow a}^{(t)}-r_i^{(t)}\\
&\approx -\Sigma_i^{(t)}\frac{h_{ai}^{*}(y_a-Z_{i\leftarrow a}^{(t)})}{1+V_a^{(t)}},
\label{Equ:F3}
\end{align}
where we use the approximations $V_a^{(t)}=V_{i\leftarrow a}^{(t)}+O(1/N)$ and $\Sigma_i^{(t)}=\Sigma_{i\rightarrow a}^t+O(1/N)$ to obtain $\triangle r$. Applying the fact\footnote
{
 Provided that $f(x)$ is an arbitrary bounded and non-negative function and define a distribution $\pa(x)=\frac{f(x)\mathcal{N}(x|m,v)}{\int f(x)\mathcal{N}(x|m,v) \text{d}x}$. Denote its mean and variance as $\mathbb{E}\{x\}=\int x \pa(x)\text{d}x$ and $\text{Var}\{x\}=\int (x-\mathbb{E}\{x\})^2\pa(x)\text{d}x$. We have
$\frac{\partial \int x\pa(x)\text{d}x}{\partial m}=\frac{\int x\frac{x-m}{v}f(x)\mathcal{N}(x|m,v) \text{d}x\cdot \int f(x)\mathcal{N}(x|m,v) \text{d}x}{\left[\int f(x)\mathcal{N}(x|m,v) \text{d}x\right]^2}-\frac{\int xf(x)\mathcal{N}(x|m,v)\text{d}x\cdot \int \frac{x-m}{v}f(x)\mathcal{N}(x|m,v)\text{d}x}{\left[\int f(x)\mathcal{N}(x|m,v)\text{d}x\right]^2}=\frac{\text{Var}\{x\}}{v}$.
} $\frac{\partial }{\partial r}\textsf{F}_{\beta}(x_i;r,\Sigma_i^{(t)})|_{r=r_i^{(t)}}=\frac{\beta}{\Sigma_i^{(t)}}\textsf{G}_{\beta}(x;r_i^{(t)}, \Sigma_i^{(t)})=\frac{\hat{v}_i^{(t+1)}}{\Sigma_i^{(t)}}$,  (\ref{Equ:F2}) can be simplified as
\begin{align}
\hat{x}_{i\rightarrow a}^{(t+1)}
&\approx \hat{x}_i^{(t+1)}-\frac{h_{ai}^{*}(y_a-Z_{i\leftarrow a}^{(t)})}{1+V_{a}^{(t)}}\hat{v}_i^{(t+1)}.
\label{Equ:F5}
\end{align}

Applying Taylor series to $\hat{v}_{i\rightarrow a}^{(t+1)}$ in (\ref{Equ:Fa}), we have
\begin{align}
\hat{v}_{i\rightarrow a}^{(t+1)}\approx \hat{v}_i^{(t+1)}+\triangle r\frac{\partial }{\partial r}\beta \textsf{G}_{\beta}(x_i;r_{i}^{(t)},\Sigma_{i}^{(t)}).
\label{Equ:F4}
\end{align}
Combining (\ref{Equ:F3}) with (\ref{Equ:F4}) into (\ref{Equ:Va}) obtains
\begin{align}
\nonumber
V_a^{(t)}&=\sum_{i=1}^N |h_{ai}|^2\left(\hat{v}_i^{(t)}-\Sigma_i^{(t)}\frac{h_{ai}^{*}(y_a-Z_{i\leftarrow a}^{(t)})}{1+V_{a}^{(t)}}\right.\\
\nonumber
&\qquad \times \left. \frac{\partial }{\partial r}\beta\textsf{G}_{\beta}(x_i;r,\Sigma_{i}^{(t)})\right)\\
\nonumber
&\approx \sum_{i=1}^N |h_{ai}|^2\hat{v}_i^{(t)}-\sum_{i=1}^N\frac{|h_{ai}|^3(y_a-Z_{i\leftarrow a}^{(t)})}{\sum_{a=1}^M |h_{ai}|^2}\\
\nonumber
&\qquad \times \frac{\partial }{\partial r}\beta\textsf{G}_{\beta}(x_i;r,\Sigma_{i}^{(t)})\\
\nonumber
&= \sum_{i=1}^N |h_{ai}|^2\hat{v}_i^{(t)}+O(1/\sqrt{N})\\
&\approx \sum_{i=1}^N |h_{ai}|^2\hat{v}_i^{(t)}.
\end{align}
Substituting (\ref{Equ:F5}) into (\ref{Equ:Za}) gets
\begin{align}
\nonumber
Z_a^{(t)}&\approx \sum_{i=1}^N h_{ai}\hat{x}_i^{(t)}-\sum_{i=1}^N\frac{ |h_{ai}|^2(y_a-Z_{i\leftarrow a}^{(t-1)})}{1+V_{a}^{(t-1)}}\hat{v}_i^{(t)}\\
\nonumber
&= \sum_{i=1}^N h_{ai}\hat{x}_i^{(t)}-\sum_{i=1}^N\frac{ |h_{ai}|^2\hat{v}_i^{(t)}(y_a-Z_a^{(t-1)}+h_{ai}\hat{x}_i^{(t-1)})}{1+V_{a}^{(t-1)}}\\
&\approx \sum_{i=1}^N h_{ai}\hat{x}_i^{(t)}-\frac{V_a^{(t)}(y_a-Z_a^{(t-1)})}{1+V_a^{(t-1)}}.
\end{align}
Inserting (\ref{Equ:F5}) into (\ref{Equ:F7}) yields
\begin{align}
\nonumber
r_i^{(t)}
&\approx \Sigma_{i}^{(t)}\sum_{a=1}^M\frac{h_{ai}^{*}(y_a-Z_a^{(t)}+h_{ai}\hat{x}_i^{(t)})}{1+V_{a}^{(t)}}\\
&= \hat{x}_i^{(t)}+\Sigma_i^{(t)}\sum_{a=1}^M\frac{h_{ai}^{*}(y_a-Z_a^{(t)})}{1+V_a^{(t)}}.
\end{align}
Up to now, the derivation of AMP for LASSO is complete. The AMP algorithm is shown in Algorithm \ref{Algorithm:AMP}.

\begin{algorithm}[!t]
\caption{AMP for LASSO}
\label{Algorithm:AMP}
{
\begingroup
\textbf{1. Input:} $\bs{y}$, $\bs{H}$.\\
\textbf{2. Initialization:} $\hat{x}_i^{(1)}=0$, $\hat{v}_i^{(1)}=1$, $Z_a^{(0)}=y_a$.\\
\textbf{3. Output:} $\hat{\bs{x}}^{(T)}$.\\
\textbf{4. Iteration:} \\
\For{$t=1,\cdots, T$}
{
 \setlength\abovedisplayskip{0pt}
 \setlength\belowdisplayskip{0pt}
 \begin{subequations}
 \begin{align}
 V_a^{(t)}&=\sum_{i=1}^N |h_{ai}|^2\hat{v}_i^{(t)}\\
 Z_a^{(t)}&=\sum_{i=1}^Nh_{ai}\hat{x}_i^{(t)}-\frac{V_a^{(t)}(y_a-Z_a^{(t-1)})}{1+V_a^{(t-1)}}\\
 \Sigma_i^{(t)}&=\left(\sum_{a=1}^M\frac{|h_{ai}|^2}{1+V_a^{(t)}}\right)^{-1}\\
 r_i^{(t)}&= \hat{x}_i^{(t)}+\Sigma_i^{(t)}\sum_{a=1}^M\frac{h_{ai}^{*}(y_a-Z_a^{(t)})}{1+V_a^{(t)}}\\
 \hat{x}_{i}^{(t+1)}&=\textsf{F}_{\beta}(x_i;r_i^{(t)},\Sigma_i^{(t)})\\
 \hat{v}_{i}^{(t+1)}&=\beta \textsf{G}_{\beta}(x_i;r_i^{(t)},\Sigma_i^{(t)})
 \end{align}
 \end{subequations}
}
\endgroup
}
\end{algorithm}

To in line with Donoho's AMP, we still need to carry out the following simplifications using the fact $|h_{ai}|^2=O(1/M)$
\begin{subequations}
\begin{align}
V_a^{(t)}&=\frac{1}{M}\sum_{i=1}^N \hat{v}_i^t\overset{\triangle}{=}V^{(t)},\\
Z_a^{(t)}&=\sum_{i=1}^N h_{ai}\hat{x}_i^{(t)}-\frac{V^{(t)}(y_a-Z_a^{(t-1)})}{1+V^{(t-1)}},\\
\Sigma_i^{(t)}&=1+V^{(t)}\overset{\triangle}{=}\Sigma^{(t)},\\
r_i^{(t)}&=\hat{x}_i^{(t)}+\sum_{a=1}^M h_{ai}^{*}(y_a-Z_a^{(t)}),\\
\hat{x}_i^{(t+1)}&=\textsf{F}_{\beta}(x_i;r_i^{(t)}, \Sigma^{(t)}),\\
\hat{v}_i^{(t+1)}&=\Sigma^{(t)}\textsf{F}_{\beta}'(x_i;r_i^{(t)}, \Sigma^{(t)}),
\end{align}
\end{subequations}
\!\!where $\textsf{F}_{\beta}'(x_i;r_i^{(t)}, \Sigma^{(t)})$ is the partial derivation of $\textsf{F}_{\beta}(x_i;r_i^{(t)}, \Sigma^{(t)})$ w.r.t. $r_i^{(t)}$.

Defining $\bs{z}^{(t)}=\bs{y}-\bs{Z}^{(t)}$ with $\bs{Z}^{(t)}\overset{\triangle}{=}\{Z_a^{(t)},\forall a\}$, we have
\begin{subequations}
\begin{align}
\nonumber
\bs{z}^{(t)}&=\bs{y}-\bs{H}\hat{\bs{x}}^{(t)}\\
&\ \ +\frac{1}{\alpha}\bs{z}^{(t-1)}\left<\textsf{F}_{\beta}'(\bs{x}; \hat{\bs{x}}^{(t-1)}+\bs{H}^{\text{T}}\bs{z}^{(t-1)}, \Sigma^{(t-1)})\right>,\\
\hat{\bs{x}}^{(t+1)}&=\textsf{F}_{\beta}(\bs{x};\hat{\bs{x}}^{(t)}+\bs{H}^{\text{T}}\bs{z}^{(t)}), \Sigma^{(t)}),\\
\Sigma^{(t+1)}&=\Sigma^{(t)}\left<\textsf{F}_{\beta}'(\bs{x};\hat{\bs{x}}^{(t)}+\bs{H}^{\text{T}}\bs{z}^{(t)}, \Sigma^{(t)})\right>.
\end{align}
\end{subequations}
In large $\beta$, by Laplace method of integration we have
\begin{align}
\nonumber
&\lim_{\beta\rightarrow \infty}\textsf{F}_{\beta}(x_i;r_i^{(t)}, \Sigma^{(t)})\\
\nonumber
&=
\lim_{\beta\rightarrow \infty}\int x_i \frac{1}{Z^{\text{pos}}}\exp \left[-\beta\left(\lambda |x_i|+\frac{1}{2\Sigma^{(t)}}(x_i-r_i^{(t)})\right)\right]\text{d}x_i\\
&=\underset{x_i}{\arg \min} \ \frac{1}{2\Sigma^{(t)}}(x_i-r_i^{(t)})^2+\lambda |x_i|.
\end{align}
Similar to (\ref{Equ:B1})-(\ref{Equ:B2}), we get
\begin{align}
\lim_{\beta\rightarrow \infty}\textsf{F}_{\beta}(x_i;r_i^{(t)}, \Sigma^t)&=\text{sign}(r_i^{(t)})\max(|r_i^{(t)}|, \lambda \Sigma^{(t)}),\\
\lim_{\beta\rightarrow \infty}\textsf{F}_{\beta}'(x_i;r_i^t, \Sigma^{(t)})&=
\begin{cases}
1 &|r_i^{(t)}|\geq\Sigma^{(t)}\\
0 & \text{otherwise}
\end{cases}.
\end{align}
Defining $\eta(r, \gamma)=\text{sign}(r)\max(|r|, \gamma)$ and $\hat{\tau}^{(t)}=\lambda V^{(t)}$, we have
\begin{subequations}
\begin{align}
\nonumber
\bs{z}^{(t)}&=\bs{y}-\bs{H}\hat{\bs{x}}^{(t)}\\
&\quad +\frac{1}{\alpha}\bs{z}^{(t-1)}\left<\eta'(\hat{\bs{x}}^{(t-1)}+\bs{H}^{\text{T}}\bs{z}^{(t-1)}, \lambda+\hat{\tau}^{(t-1)})\right>,\\
\hat{\bs{x}}^{(t+1)}&=\eta(\hat{\bs{x}}^{(t)}+\bs{H}^{\text{T}}\bs{z}^{(t)}, \lambda+\hat{\tau}^{(t)}),\\
\hat{\tau}^{(t+1)}&=\frac{\lambda+\hat{\tau}^{(t)}}{\alpha}\left<\eta'(\hat{\bs{x}}^{(t)}+\bs{H}^{\text{T}}\bs{z}^{(t)}, \lambda+\hat{\tau}^{(t)})\right>.
\end{align}
\label{Alg:AMP2}
\end{subequations}
\!\!\!By abusing $\eta$, we get the original AMP (\ref{Alg:AMP}) for LASSO inference problem.

\begin{algorithm}[!t]
\caption{Bayes-Optimal AMP}
\label{Algorithm:BAMP}
{
\begingroup
\textbf{1. Input:} $\bs{y}$, $\bs{H}$, $\sigma_w^2$, $\pa(\bs{x})$. \\
\textbf{2. Initialization:} $\hat{x}_i^{(1)}=0$, $\hat{v}_i^{(1)}=1$, $Z_a^{(0)}=y_a$.\\
\textbf{3. Output:} $\hat{\bs{x}}^{(T)}$.\\
\textbf{4. Iteration:} \\
\For{$t=1,\cdots, T$}
{
 \setlength\abovedisplayskip{0pt}
 \setlength\belowdisplayskip{0pt}
 \begin{subequations}
 \begin{align}
 V_a^{(t)}&=\sum_{i=1}^N |h_{ai}|^2\hat{v}_i^{(t)}\\
 Z_a^{(t)}&=\sum_{i=1}^Nh_{ai}\hat{x}_i^{(t)}-\frac{V_a^{(t)}(y_a-Z_a^{(t-1)})}{\sigma_w^2+V_a^{(t-1)}}\\
 \Sigma_i^{(t)}&=\left(\sum_{a=1}^M\frac{|h_{ai}|^2}{\sigma_w^2+V_a^{(t)}}\right)^{-1}\\
 r_i^{(t)}&= \hat{x}_i^{(t)}+\Sigma_i^{(t)}\sum_{a=1}^M\frac{h_{ai}^{*}(y_a-Z_a^{(t)})}{\sigma_w^2+V_a^{(t)}}\\
 \hat{x}_{i}^{(t+1)}&=\mathbb{E}\{x_i|r_i^{(t)},\Sigma_i^{(t)}\}
 \label{Equ:AMP5}\\
 \hat{v}_{i}^{(t+1)}&=\text{Var}\{x_i|r_i^{(t)}, \Sigma_i^{(t)}\}
 \label{Equ:AMP6}
 \end{align}
 \end{subequations}
}
\endgroup
}
\end{algorithm}

\subsection{Bayes-optimal AMP}
In LASSO inference problem, both the prior and likelihood are unknown. However, in the Bayes-optimal setting, where both prior and likelihood function are perfectly given, the MMSE estimator can achieve Bayes-optimal error. Actually, this situation is common in communications. In those cases, it is assumed that each element of $\bs{x}$ follows IID distribution $\pa_{\textsf{X}}$. The joint distribution is then represented as
\begin{align}
\nonumber
\pa(\bs{x},\bs{y})
&=\pa(\bs{y}|\bs{x})\pa(\bs{x})\\
&=\prod_{a=1}^M\pa(y_a|\bs{x})\prod_{i=1}^N \pa_{\textsf{X}}(x_i).
\end{align}
Similar to the derivation of AMP for LASSO, we get the Bayes-optimal AMP as depicted in Algorithm~\ref{Algorithm:BAMP}, where the expectation in (\ref{Equ:AMP5}) and (\ref{Equ:AMP6}) is taken over
\begin{align}
\hat{\pa}^{(t)}(x_i|\bs{y})=\frac{\pa_{\textsf{X}}(x_i)\mathcal{N}(x_i|r_i^{(t)}, \Sigma_i^{(t)})}{\int \pa_{\textsf{X}}(x)\mathcal{N}(x|r_i^{(t)}, \Sigma_i^{(t)}) \text{d}x}.
\end{align}
This form of AMP is widely applied to many engineering regions. We call it as Bayes-optimal AMP since (1) this algorithm is based on Bayes-optimal setting;  (2) the SE of this algorithm perfectly matches the fixed point of the exact MMSE estimator predicted by replica method. Similar to AMP for LASSO, the form of Bayes-optimal AMP can also be written as (\ref{Alg:AMP2}) with $\eta(\cdot)$ being MMSE denoiser.

\subsection{State Evolution}
In this subsection, we only give a sketch of proving AMP's SE in \cite{bayati2011dynamics}. Let's introduce the following general iterations.
\begin{subequations}
\begin{align}
\bs{h}^{(t+1)}&=\bs{H}^{\text{T}}\bs{m}^{(t)}-\xi_t\bs{q}^{(t)},\\
\bs{b}^{(t)}&=\bs{H}\bs{q}^{(t)}-\lambda_t\bs{m}^{(t-1)},
\end{align}
\label{General_AMP}
\end{subequations}
\!\!\!where $\bs{m}^{(t)}=g_{t}(\bs{b}^{(t)}, \bs{n})$, $\bs{q}^{(t)}=f_t(\bs{h}^{(t)},\bs{x})$, $\xi_t=\left<g_t'(\bs{b}^{(t)}, \bs{n})\right>$, and $\lambda_t=\frac{1}{\alpha}\left<f_t'(\bs{h}^{(t)}, \bs{x})\right>$.

Pertaining to this general iterations, the following conclusions can be established. In the large system limit, for any pseudo-Lipschitz function $\varphi:\mathbb{R}^2\mapsto\mathbb{R}$ of order $k$ and all $t\geq 0$,
\begin{subequations}
\begin{align}
\lim_{N\rightarrow \infty}\frac{1}{N}\sum_{i=1}^N\varphi(h_i^{(t+1)}, x_i)&\overset{\text{a.s.}}{=}\mathbb{E}_{\textsf{Z}, \textsf{X}}\left\{\varphi(\tau_t\textsf{Z}, \textsf{X})\right\},\\
\lim_{M\rightarrow \infty}\frac{1}{M}\sum_{i=1}^M \varphi(b_i^{(t)}, n_i)&\overset{\text{a.s.}}{=}\mathbb{E}_{\textsf{Z}, \textsf{N}}\left\{\varphi(\sigma_t\textsf{Z}, \textsf{N})\right\},
\end{align}
\label{Equ:H3}
\end{subequations}
\!\!where
\begin{align}
\tau_t^2&=\mathbb{E}\left\{g_t(\sigma_t\textsf{Z}, \textsf{N})^2\right\},
\label{Equ:H1}\\
\sigma_t^2&=\frac{1}{\alpha}\mathbb{E}\left\{f_t(\tau_{t-1}\textsf{Z}, \textsf{X})^2\right\},
\label{Equ:H2}
\end{align}
where $\textsf{N}\sim \pa_{\textsf{N}}$ and $\textsf{X}\sim \pa_{\textsf{X}}$ are independent of $\textsf{Z}\sim \mathcal{N}(0,1)$. Specially, $\sigma_0^2=\lim_{N\rightarrow \infty}\frac{1}{N\alpha}\|\bs{q}^{(0)}\|^2$.

Define
\begin{align}
g_t(\bs{b}^{(t)}, \bs{n})&=\bs{b}^{(t)}-\bs{n},
\label{Reduce-1}\\
f_t(\bs{h}^{(t)},\bs{x})&=\eta_{t-1}(\bs{x}-\bs{h}^{(t)})-\bs{x}.
\label{Reduce-2}
\end{align}
Then $\xi_t=1$ and $\lambda_t=-\frac{1}{\alpha}\left<\eta_{t-1}'(\bs{x}-\bs{h}^{(t)})\right>$.
To coincide with AMP (Donoho) in (\ref{Alg:AMP}), it implies that $\bs{x}-\bs{h}^{(t+1)}=\bs{H}^{\text{T}}\bs{z}^{(t)}+\bs{x}^{(t)}$. We thus have
\begin{subequations}
\begin{align}
\bs{h}^{(t+1)}&=\bs{x}-(\bs{H}^{\text{T}}\bs{z}^{(t)}+\bs{x}^{(t)})\\
\bs{q}^{(t)}&=\hat{\bs{x}}^{(t)}-\bs{x},\\
\bs{b}^{(t)}&=\bs{n}-\bs{z}^{(t)},\\
\bs{m}^{(t)}&=-\bs{z}^{(t)}.
\end{align}
\label{Reduce-4}
\end{subequations}
\!\!Using (\ref{Reduce-1})-(\ref{Reduce-2}) and (\ref{Reduce-4}), the general iterative equations (\ref{General_AMP}) reduce to the original AMP.

From (\ref{Equ:H1})-(\ref{Reduce-2}), we get the SE of AMP
\begin{align}
\nonumber
\tau_{t+1}^2
&=\sigma^2_w+\sigma_{t+1}^2\\
&=\sigma^2_w+\frac{1}{\alpha}\mathbb{E}\left\{(\eta_{t}(\textsf{X}+\tau_t\textsf{Z})-\textsf{X})^2\right\}.
\end{align}

\noindent For the proof of AMP's SE, we have the following remarks:
\begin{remark}
\textit{Conditional distribution}.
To prove the equations (\ref{Equ:H3}), the so-called condition technique is applied. Let's consider a linear constrain $\bs{Y}=\bs{AX}$, where $\bs{A}$ follows $\pa_{\bs{A}}(\bs{A})$. Let $G$ denote the event that $\bs{A}$ satisfies the linear constrain $\bs{Y}=\bs{AX}$. Then we say that $\bs{A}$ under $G$ is distributed as $\bs{B}$ following
\begin{align}
\pa_{\bs{A}|G}(\bs{B})=\frac{1}{Z}\pa_{\bs{A}}(\bs{B})\cdot \mathbbm{1}_{\bs{B}\in \mathcal{L}},
\end{align}
where $Z$ is normalized constant and $\mathcal{L}$ denotes the set of $\bs{A}$ that satisfies the linear constrain $\bs{Y}=\bs{AX}$. We write it as $\bs{A}|_G\overset{\text{d}}{=}\bs{B}$.\\
\textit{Gaussianity}. The equations (\ref{Equ:H3}) shows that in the large system limit, each entry of $\bs{h}^{(t+1)}$ and $\bs{b}^{(t)}$ tends to Gaussian RV. Regarding $\bs{h}^{(t)}$ and $\bs{b}^{(t)}$ as column vectors, then for $t\geq 0$, from (\ref{General_AMP}), we have
\begin{align}
\nonumber
&\underbrace{\left[\bs{h}^{(1)}+\xi_0\bs{q}^{(0)},\cdots, \bs{h}^{(t)}+\xi_{t-1}\bs{q}^{(t-1)}\right]}_{\overset{\triangle}{=}\bs{X}_t}\\
&=\bs{H}^{\text{T}}\underbrace{\left[\bs{m}^{(0)},\cdots, \bs{m}^{(t-1)}\right]}_{\overset{\triangle}{=}\bs{M}_t},\\
\nonumber
&\underbrace{\left[\bs{b}^{(0)},\bs{b}^{(1)}+\lambda_1\bs{m}^{(0)},\cdots, \bs{b}^{(t-1)}+\lambda_{t-1}\bs{m}^{(t-2)}\right]}_{\overset{\triangle}{=}\bs{Y}_t}\\
&=\bs{H}\underbrace{\left[\bs{q}^{(0)},\cdots, \bs{q}^{(t-1)}\right]}_{\overset{\triangle}{=}\bs{Q}_t}.
\end{align}
Let $G_{t_1,t_2}$ denote the event that $\bs{H}$ satisfies the linear constrains $\bs{X}_{t_1}=\bs{H}^{\text{T}}\bs{M}_{t_1}$ and $\bs{Y}_{t_2}=\bs{H}\bs{Q}_{t_2}$. Then the conditional distribution of $\bs{h}^{(t+1)}$ and $\bs{b}^{(t)}$ can be expressed as
\begin{align}
\bs{h}^{(t+1)}|_{G_{t+1,t}}&\overset{\text{d}}{=}\bs{H}|_{G_{t+1,t}}\bs{m}^{(t)}-\xi_t\bs{q}^{(t)},\\
\bs{b}^{(t)}|_{G_{t,t}}&\overset{\text{d}}{=}\bs{H}|_{G_{t,t}}\bs{q}^{(t)}-\lambda_t\bs{m}^{(t-1)}.
\end{align}
The approximated expressions are shown in \cite[Lemma 1]{bayati2011dynamics}, where $t$-iteration $\bs{h}^{(t+1)}$ (or $\bs{b}^{(t)}$) on the conditions $G_{t+1,t}$ (or $G_{t,t}$) is expressed as a combination of all preceding $\{\bs{h}^{(\tau)}, \forall \tau\leq t\}$ (or $\{\bs{b}^{(\tau)}, \tau<t\}$). The proof of Lemma 1 is rigorous since the induction on $t$ is rigorous. Be aware, during the proof of Lemma 1, the fact that $\bs{H}$ has IID Gaussian entries is applied to derive the Gaussianity of $\bs{h}^{(t+1)}$ and $\bs{b}^{(t)}$.
\end{remark}

%
%

\subsection{Numeric Simulations}
\begin{figure}[!t]
\centering
\includegraphics[width=0.43\textwidth]{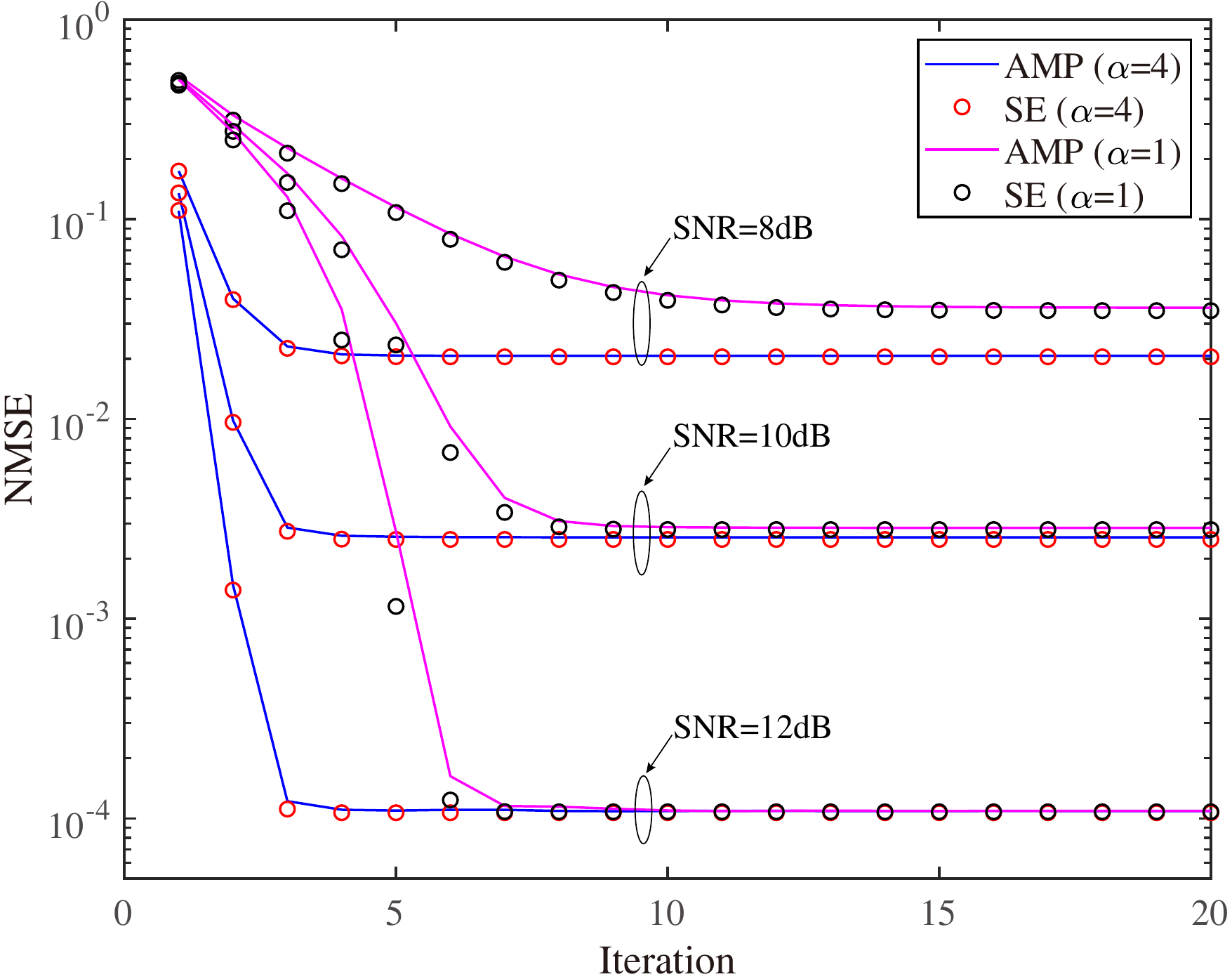}
\caption{Iterative behavior of Bayes-optimal AMP and its SE in wireless communications. $\bs{H}$ has IID Gaussian entries with zero mean and $1/M$ variance. $M=1024$, $N=\frac{M}{\alpha}$ and SNR=$1/\sigma_w^2$. The signal of interest $\bs{x}$ has IID entries from the set $\{\pm\frac{1}{\sqrt{2}}\pm\mathbb{J}\frac{1}{\sqrt{2}}\}$ with $\mathbb{J}^2=-1$.
}
\label{EXP2A}
\vspace{+0.2cm}
\includegraphics[width=0.43\textwidth]{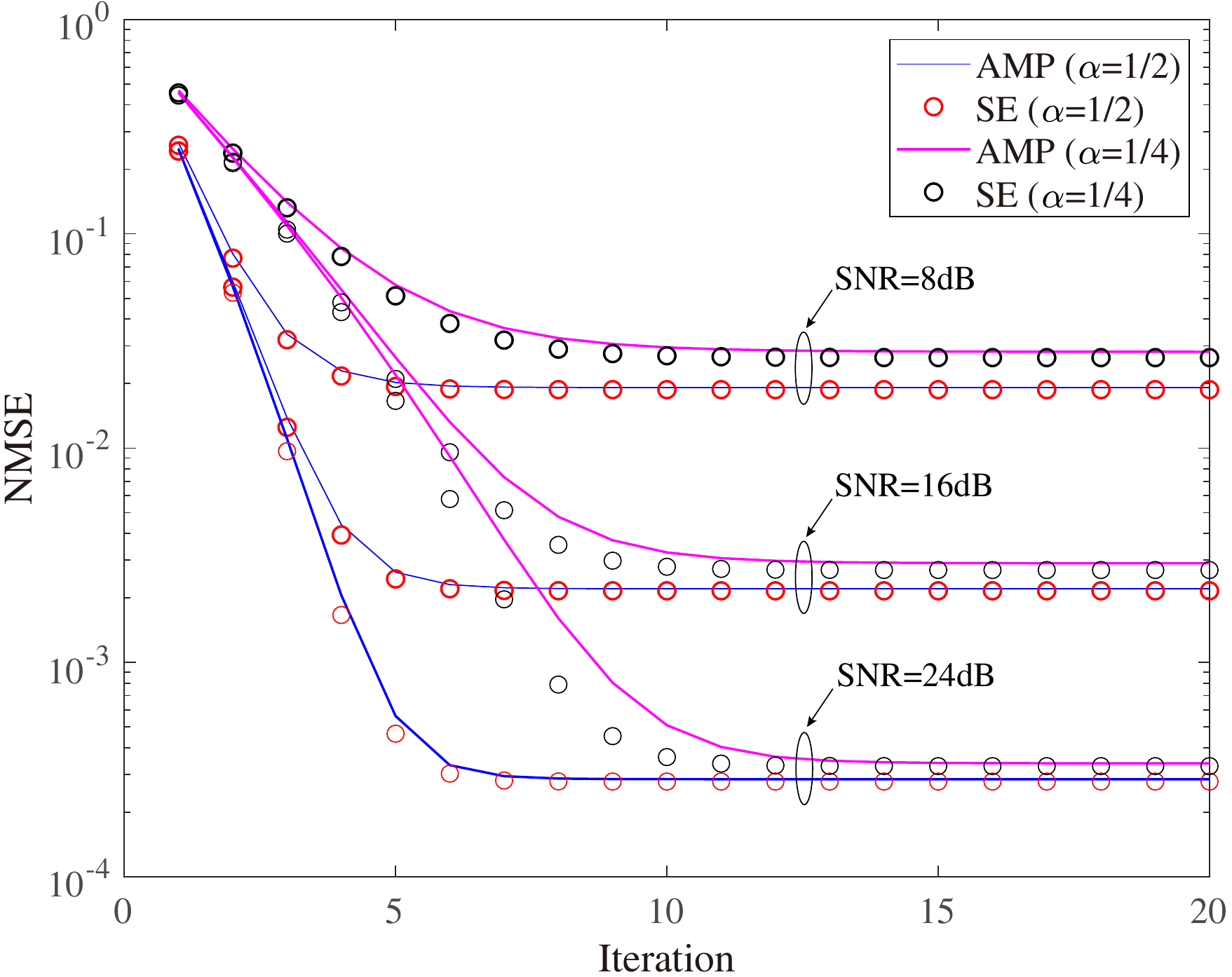}
\caption{Iterative behavior of Bayes-optimal AMP and its SE in compressed sensing.  $\bs{H}$ has IID Gaussian entries with zero mean and $1/M$ variance. $M=\alpha N$, $N=1024$ and SNR=$1/\sigma_w^2$. The signal of interest $\bs{x}$ has IID entries following $\mathcal{BG}(0,0.05)$.
}
\label{EXP2B}
\end{figure}

In Fig.~\ref{EXP2A}, we show the comparison of Bayes-optimal AMP and its SE in the application of wireless communications. As can be seen from Fig.~\ref{EXP2A}, firstly, AMP matches the SE curve very well; secondly, the performance of AMP becomes better as SNR increases; finally, in small SNR, the measurement ratio $\alpha$ has the effect on convergence speed and fixed point while in large SNR, the effect of $\alpha$ on fixed point can be ignored in the case of QPSK prior. Specially, as SNR=12dB, the curves of $\alpha=1$ and $\alpha=4$ converge to the same fixed point almost sure.

In Fig~\ref{EXP2B}, we show the comparison of Bayes-optimal AMP and its SE in compressed sense. As can be observed from Fig.~\ref{EXP2B}, AMP matches the SE curves in all settings. We also see that similar to application in wireless communications, as SNR increases, the gap between the difference measurement ratios will be decreased. Besides, as measurement ratio increases, the convergence speed of AMP will be faster.

\section{From AMP to OAMP}
Although AMP can achieve the Bayes-optimal MSE performance in IID sub-Gaussian region, the AMP algorithm may fail to converge when $\bs{H}$ is ill-conditioned (e.g. large conditional number, non-zero mean). To extend the scope of AMP to more general random matrices (unitrarily-invariant matrix\footnote{
We say $\bs{A}=\bs{U\Sigma V}^{\text{T}}$ is unitarily-invariant if $\bs{U}$, $\bs{V}$, and $\bs{\Sigma}$ are mutually independent, and $\bs{U}$, $\bs{V}$ are Haar-distributed.
}), a modified AMP algorithm termed OAMP \cite{ma2017orthogonal} was proposed. Different from AMP, the denoiser of OAMP is divergence-free so that the Onsager term vanishes and the LMMSE de-correlated matrix is applied to ensure the orthogonality\footnote{
Given two random variables $\textsf{X}$, $\textsf{Y}$, we say $\textsf{X}$ is orthogonal to $\textsf{Y}$ if $\mathbb{E}\{\textsf{XY}\}=0$. Provided that $\bs{x}\in \mathbb{R}^N$ and $\bs{y}\in \mathbb{R}^N$ are generated by $\textsf{X}$ and $\textsf{Y}$, respectively, then $\frac{1}{N}\bs{x}^{\text{T}}\bs{y}=\frac{1}{N}\sum_{i=1}^Nx_iy_i\overset{\text{a.s.}}{=}\mathbb{E}\{\textsf{XY}\}=0$.
} of input and output errors of denoiser.

\subsection{Orthogonality of input and output errors}
Let's consider the following general iterations containing a linear estimation (LE) and a nonlinear estimation (NLE):
\begin{subequations}
\begin{align}
&\text{LE}: \qquad \ \bs{r}^{(t)}=\hat{\bs{x}}^{(t)}+\bs{W}_t(\bs{y}-\bs{H}\hat{\bs{x}}^{(t)})+\bs{r}^{(t)}_{\text{Onsager}},\\
&\text{NLE}: \ \hat{\bs{x}}^{(t+1)}=\tilde{\eta}_{t}(\bs{r}^{(t)}).
\end{align}
\label{GeneralAMP}
\end{subequations}
where $\bs{W}_t$ is a linear transform matrix that maps residual error $\bs{y}-\bs{H}\hat{\bs{x}}^{(t)}$ onto $\mathbb{R}^N$, $\bs{r}_{\text{Onsager}}^{(t)}$ is the Onsager term, and $\tilde{\eta}_t$ is the denoiser. Specially, as $\bs{r}_{\text{Onsager}}^{(t)}=\frac{\bs{r}^{(t-1)}-\hat{\bs{x}}^{(t-1)}}{\alpha}\left<\eta_{t-1}'(\bs{r}^{(t-1)})\right>$, $\bs{W}_t=\bs{H}^{\text{T}}$, and $\tilde{\eta}_t(\bs{r}^{(t)})=\eta_t(\bs{r}^{(t)})$, the above general iterations reduce to Donoho's AMP. In AMP algorithm, the Onsager term $\bs{r}_\text{Onsager}^{(t)}$ ensures the Gaussianity of input signal $\bs{r}^{(t)}$ and AMP can achieve Bayes-optimal performance in IID sub-Gaussian random measurement matrix. A significant disadvantage of AMP is that AMP may diverge when the random measurement matrix is beyond IID sub-Gaussian. To extend the scope of AMP  to more general case, \cite{ma2017orthogonal} proposed a modified AMP algorithm called OAMP.

The main ideal of OAMP is to design a linear transform matrix  $\bs{W}_t$ and denoiser $\tilde{\eta}_t$ so that
\begin{itemize}
\item \textbf{Divergence-free}\footnote{We say $\eta: \mathbb{R}\mapsto \mathbb{R}$ is divergence-free if $\mathbb{E}\{\eta'(R)\}=0$.}.
 The modified algorithm does not dependent on the Onsager term so that the Onsager term vanishes;
\item \textbf{Orthogonality}.
 The modified algorithm maintains the orthogonality of the input and output errors of denoiser $\tilde{\eta}_t(\cdot)$.
\end{itemize}
 For the first issue, a divergence-free denoiser $\tilde{\eta}_t$ can be constructed as
\begin{align}
\tilde{\eta}_t(\bs{r}^{(t)})=C\left[\eta_t(\bs{r}^{(t)})-\bs{r}^{(t)}\left<\eta_t'(\bs{r}^{(t)})\right>\right],
\label{Equ:DF}
\end{align}
where $\eta_t(\cdot)$ can be an arbitrary pseudo-Lipschitz function and $C$ is a constant. In this case, we have $\left<\tilde{\eta}_t'(\bs{r}^{(t)})\right>=0$.

For convenience, we define the input and output errors
\begin{align}
\bs{q}^{(t)}&=\hat{\bs{x}}^{(t)}-\bs{x},\\
\bs{h}^{(t)}&=\bs{r}^{(t)}-\bs{x}.
\end{align}
Substituting the system model $\bs{y}=\bs{Hx}+\bs{n}$ and (\ref{GeneralAMP}) into equations above, we have
\begin{subequations}
\begin{align}
&\text{LE}: \qquad  \bs{h}^{(t)}=(\bs{I}-\bs{W}_{t}\bs{H})\bs{q}^{(t)}+\bs{W}_t\bs{n},\\
&\text{NLE}: \ \bs{q}^{(t+1)}=\tilde{\eta}_t(\bs{x}+\bs{h}^{(t)})-\bs{x}.
\end{align}
\label{OAMP_General}
\end{subequations}
\!\!Also, we define error-related parameters
\begin{align}
\hat{v}^{(t)}&=\lim_{N\rightarrow \infty}\frac{1}{N}\|\hat{\bs{q}}^{(t)}\|_2^2, \quad \tau_t^2=\lim_{N\rightarrow \infty}\frac{1}{N}\|\bs{h}^{(t)}\|_2^2.
\end{align}

Similar to AMP, we assume that the following assumptions hold
\begin{itemize}
\item \textbf{Assumption 1}: the input error $\bs{h}^{(t)}$ consists of IID zero-mean Gaussian entries independent of $\bs{x}$, i.e., $\textsf{R}^{(t)}=\textsf{X}+\tau_t\textsf{Z}$ with $\textsf{Z}$ being a standard Gaussian RV.
\item \textbf{Assumption 2}: the output error $\bs{q}^{(t+1)}$ consists of IID entries independent of $\bs{H}$ and noise $\bs{n}$.
\end{itemize}
We will show that based on the assumptions above, the de-correlated matrix $\bs{W}_t$ and divergence-free imply the orthogonality between input error $\bs{h}^{(t)}$ and output error $\bs{q}^{(t+1)}$. We say LE is de-correlated one if $\text{Tr}(\bs{I}-\bs{W}_t\bs{H})=0$, which implies
\begin{align}
\bs{W}_t=\frac{N}{\text{Tr}(\hat{\bs{W}}_t\bs{H})}\hat{\bs{W}}_t,
\end{align}
where $\hat{\bs{W}}_t$ can be chosen from:
\begin{align}
\hat{\bs{W}}_t=
&
\begin{cases}
\bs{H}^{\text{T}}   &\text{matched filter (MF)}\\
\hat{\bs{W}}_t^{\text{pinv}}  &\text{pseudo-inverse}\\
\bs{H}^{\text{T}}\left(\bs{HH}^{\text{T}}+\frac{\sigma^2}{\hat{v}^{(t)}}\bs{I}\right)^{-1} &\text{LMMSE}
\end{cases}
\end{align}
where $\hat{\bs{W}}_t^{\text{pinv}}=\bs{H}^{\text{T}}(\bs{HH}^{\text{T}})^{-1}$ for $M<N$ and $\hat{\bs{W}}_t^{\text{pinv}}=(\bs{H}^{\text{T}}\bs{H})^{-1}\bs{H}^{\text{T}}$ for $M\geq N$. Considering LMMSE de-correlated matrix, one should need to determine $\hat{v}^{(t)}$. We consider $\eta_t$ as MMSE denoiser and denote it as $\eta_{t}^{\text{mmse}}$ to distinguish $\tilde{\eta}_t$. Based on Assumption 1, from (\ref{Equ:DF}), we have
\begin{align}
\tilde{\eta}_t(\bs{r}^{(t)})&=C\hat{v}_{\text{mmse}}^{(t)}\left(\frac{\eta_{t}^{\text{mmse}}(\bs{r}^{(t)})}{\hat{v}^{(t)}_{\text{mmse}}}-\frac{\bs{r}^{(t)}}{\tau_t^2}\right),
\label{Equ:tilde_eta}
\end{align}
where the relation $\left<{\eta^{\text{mmse}}_t}'(\bs{r}^{(t)})\right>=\frac{\hat{v}_{\text{mmse}}^{(t)}}{\tau_t^2}$ is applied. The MMSE estimator and its variance are defined as
\begin{align}
\eta_t^{\text{mmse}}(r_i^{(t)})&=\mathbb{E}\{x_i|r_i^{(t)}=x+\tau_tz\},\\
\hat{v}_{\text{mmse}}^{(t)}&=\frac{1}{N}\sum_{i=1}^N\text{Var}\{x_i|r_i^{(t)}=x_i+\tau_tz\},
\end{align}
where the expectation is taken over $\frac{\pa_{\textsf{X}}(x_i)\mathcal{N}(x_i|r_i^{(t)}, \tau_t^2)}{\int \pa_{\textsf{X}}(x)\mathcal{N}(x|r_i^{(t)}, \tau_t^2) \text{d}x}$.

Then
\begin{align}
\nonumber
\!\!\!\!\!\hat{v}^{(t)}
&=\lim_{N\rightarrow \infty}\frac{1}{N}\|\bs{q}^{(t)}\|^2\\
\nonumber
&\overset{\text{a.s.}}{=}\mathbb{E}_{\textsf{Z}, \textsf{X}}\left\{\left(C\hat{v}_{\text{mmse}}^{(t)}\left(\frac{\eta_{t}^{\text{mmse}}(\textsf{X}+\tau_t\textsf{Z})}{\hat{v}^{(t)}_{\text{mmse}}}-\frac{\textsf{X}+\tau_t\textsf{Z}}{\tau_t^2}\right)-\textsf{X}\right)^2\right\}\\
&=\mathbb{E}_{\textsf{Z}, \textsf{X}}\left\{\left(C\eta_t^{\text{mmse}}(\textsf{X}+\tau_t\textsf{Z})-\frac{C\hat{v}_{\text{mmse}}^{(t)}+\tau_t^2}{\tau_t^2}\textsf{X}-\frac{C\hat{v}_{\text{mmse}}^{(t)}}{\tau_t}\textsf{Z}\right)^2\right\}
\end{align}
The coefficients of $\eta_t^{\text{mmse}}$ and $\textsf{X}$ should be equal, i.e., $C=\frac{C\hat{v}_{\text{mmse}}^{(t)}+\tau_t^2}{\tau_t^2}$, and it leads to
\begin{align}
C=\frac{\tau_t^2}{\tau_t^2-\hat{v}_{\text{mmse}}^{(t)}}.
\label{Equ:C}
\end{align}
Substituting this fact into $\hat{v}^{(t)}$ obtains
\begin{align}
\nonumber
\hat{v}^{(t)}&=\mathbb{E}_{\textsf{Z}, \textsf{X}}\left\{\left[\frac{\tau_t^2}{\tau_t^2-\hat{v}_{\text{mmse}}^{(t)}}\left(\eta_t^{\text{mmse}}(\textsf{X}+\tau_t\textsf{Z})-\textsf{X}\right)
\right.\right.\\
\nonumber
&\qquad \qquad \left.\left.-\frac{\tau_t\hat{v}_{\text{mmse}}^{(t)}}{\tau_t^2-\hat{v}_{\text{mmse}}^{(t)}}\textsf{Z}\right]^2\right\}\\
\nonumber
&=\left(\frac{\tau_t^2}{\tau_t^2-\hat{v}_{\text{mmse}}^{(t)}}\right)^2\hat{v}_{\text{mmse}}^{(t)}+\left(\frac{\tau_t\hat{v}_{\text{mmse}}^{(t)}}{\tau_t^2-\hat{v}_{\text{mmse}}^{(t)}}\right)^2\\
&=\left(\frac{1}{\hat{v}_{\text{mmse}}^{(t)}}-\frac{1}{\tau_t^2}\right)^{-1},
\end{align}
where the facts $\hat{v}_{\text{mmse}}^{(t)}\overset{\text{a.s.}}{=}\mathbb{E}_{\textsf{X}, \textsf{Z}}\{(\eta_t^{\text{mmse}}(\textsf{X}+\tau_t\textsf{Z})-\textsf{X})^2\}$ and $(\eta_t^{\text{mmse}}(\textsf{X}+\tau_t\textsf{Z})-\textsf{X})$ independent of $\textsf{Z}$ are applied.

Inserting (\ref{Equ:C}) into (\ref{Equ:tilde_eta}) obtains
\begin{align}
\tilde{\eta}_t(\bs{r}^{(t)})=\hat{v}^{(t)}\left(\frac{\eta_{t}^{\text{mmse}}(\bs{r}^{(t)})}{\hat{v}^{(t)}_{\text{mmse}}}-\frac{\bs{r}^{(t)}}{\tau_t^2}\right).
\end{align}
Be aware, there is an unknown noise-related parameter $\tau_t^2$, which is expressed as
\begin{align}
\nonumber
\tau_t^2
&=\lim_{N\rightarrow \infty}\frac{1}{N}\|\bs{h}^{(t)}\|^2\\
\nonumber
&=\frac{1}{N}\text{Tr}\left((\bs{I}-\bs{W}_t\bs{H})(\bs{I}-\bs{W}_t\bs{H})^{\text{T}}\right)\hat{v}^{(t)}\\
\nonumber
&\qquad +\frac{1}{N}\text{Tr}(\bs{W}_t\bs{W}_t^{\text{T}})\sigma_w^2\\
\nonumber
&\overset{(a)}{=}\hat{v}^{(t)}\left[\frac{N\text{Tr}(\hat{\bs{W}}_t\bs{HH}^{\text{T}}\hat{\bs{W}}_t^{\text{T}})}{\text{Tr}^2\left(\hat{\bs{W}}_t\bs{H}\right)}-\right]+\frac{N\text{Tr}(\hat{\bs{W}}_t\hat{\bs{W}}_t^{\text{T}})}{\text{Tr}^2(\hat{\bs{W}}_t\bs{H})}\sigma_w^2\\
\nonumber
&=\hat{v}^{(t)}\left[\frac{N\text{Tr}(\hat{\bs{W}}_t(\bs{HH}^{\text{T}}+\frac{\sigma_w^2}{\hat{v}^{(t)}}\bs{I})\hat{\bs{W}}_t^{\text{T}})}{\text{Tr}^2(\hat{\bs{W}}_t\bs{H})}-1\right]\\
&\overset{(b)}{=}\hat{v}^{(t)}\left(\frac{N}{\text{Tr}(\hat{\bs{W}}_t\bs{H})}-1\right),
\end{align}
where the fact $\bs{W}_t=\frac{N}{\text{Tr}(\hat{\bs{W}}_t\bs{A})}\hat{\bs{W}}_t$ is used to obtain $(a)$ and the LMMSE de-correlated matrix $\hat{\bs{W}}_t=\bs{H}^{\text{T}}\left(\bs{HH}^{\text{T}}+\frac{\sigma^2}{\hat{v}^{(t)}}\bs{I}\right)^{-1}$ is applied to obtain $(b)$. This completes the derivation of orthogonal AMP and we post OAMP algorithm in Algorithm \ref{Algorithm:OAMP}. Be aware, we here use LMMSE de-correlated $\bs{W}_t=\left(\bs{H}^{\text{T}}\bs{H}+\frac{\sigma^2}{\hat{v}^{(t)}}\bs{I}\right)^{-1}\bs{H}^{\text{T}}$ and it can be verified that this form is equal to $\bs{W}_t=\bs{H}^{\text{T}}\left(\bs{HH}^{\text{T}}+\frac{\sigma^2}{\hat{v}^{(t)}}\bs{I}\right)^{-1}$ via SVD.

\begin{algorithm}[!t]
\caption{Bayes-Optimal Orthogonal AMP}
\label{Algorithm:OAMP}
{
\begingroup
\textbf{1. Input:} $\bs{y}$, $\bs{H}$, $\sigma_w^2$, and $\pa(\bs{x})$.\\
\textbf{1.Initialization:} $\hat{x}_i^{(1)}=0$, $\hat{v}_i^{(1)}=1$, $Z_a^{(0)}=y_a$.\\
\textbf{2.Output:} $\hat{\bs{x}}^{(T)}$.\\
\textbf{3.Iteration:} \\
\For{$t=1,\cdots, T$}
{
 \setlength\abovedisplayskip{0pt}
 \setlength\belowdisplayskip{0pt}
 \begin{subequations}
\begin{align}
\hat{\bs{W}}_t&=\left(\bs{H}^{\text{T}}\bs{H}+\frac{\sigma_w^2}{\hat{v}^{(t)}}\bs{I}\right)^{-1}\bs{H}^{\text{T}}
\label{OAMP-R1}\\
\tau_t^2&=\hat{v}^{(t)}\left(\frac{N}{\text{Tr}(\hat{\bs{W}}_t\bs{H})}-1\right)
\label{OAMP-R2}\\
\bs{r}^{(t)}&=\hat{\bs{x}}^{(t)}+\frac{N}{\text{Tr}(\hat{\bs{W}}_t\bs{H})}\hat{\bs{W}}_t(\bs{y}-\bs{H}\hat{\bs{x}}^{(t)})
\label{OAMP-R3}\\
\hat{\bs{x}}_{\text{mmse}}^{(t+1)}&=\eta_t^{\text{mmse}}(\bs{r}^{(t)}, \tau_t)
\label{OAMP-R4}\\
\hat{v}^{(t)}_{\text{mmse}}&=\frac{1}{N}\sum_{i=1}^N\text{Var}\{x_i|r_i^{(t)}, \tau_t\}
\label{OAMP-R5}\\
\hat{v}^{(t+1)}&=\left(\frac{1}{\hat{v}^{(t+1)}_{\text{mmse}}}-\frac{1}{\tau_t^2}\right)^{-1}
\label{OAMP-R6}\\
\hat{\bs{x}}^{(t+1)}&=\hat{v}^{(t+1)}\left(\frac{\hat{\bs{x}}^{(t+1)}_{\text{mmse}}}{\hat{v}^{(t+1)}_{\text{mmse}}}-\frac{\bs{r}^{(t)}}{\tau_t^2}\right)
\label{OAMP-R7}
\end{align}
\end{subequations}
}
\endgroup
}
\end{algorithm}

The below is to prove orthogonality of input and output errors.  Define $\bs{B}_t=\bs{I}-\bs{W}_t\bs{H}$, we have
\begin{align}
\nonumber
\mathbb{E}\{\bs{h}^{(t)}(\bs{q}^{(t)})^{\text{T}}\}
&=\mathbb{E}\{\bs{B}_t\}\mathbb{E}\{\bs{q}^{(t)}(\bs{q}^{(t)})^{\text{T}}\}\\
\nonumber
&\quad +\mathbb{E}\{\bs{W}_t\}\mathbb{E}\{\bs{n}(\bs{q}^{(t)})^{\text{T}}\}\\
&=\mathbb{E}\{\bs{B}_t\}\mathbb{E}\{\bs{q}^{(t)}(\bs{q}^{(t)})^{\text{T}}\},
\end{align}
where the last equation holds by Assumption 2. By the SVDs $\bs{H}=\bs{U}\bs{\Sigma}\bs{V}^{\text{T}}$ and $\bs{W}_t=\bs{V}\bs{G}_t\bs{U}^{\text{T}}$, we have
\begin{align}
\mathbb{E}\{(\bs{B}_t)_{ij}\}=\sum_{m=1}^M\mathbb{E}\{V_{im}V_{jm}\}(1-g_m\sigma_m),
\end{align}
where $g_m$ is $m$-th element of $\bs{G}_t$ and $\sigma_m$ is $m$-th element of $\bs{\Sigma}$.
Since $\bs{V}$ is Haar distribution, we have
\begin{align}
\mathbb{E}\{(\bs{B}_t)_{ij}\}=&
\begin{cases}
0  &i\ne j\\
\frac{\text{Tr}(\bs{B}_t)}{N} &i=j
\end{cases}.
\end{align}
Since $\text{Tr}(\bs{B}_t)=0$, we then have $\mathbb{E}\{\bs{B}_t\}=\bs{0}$ and further
\begin{align}
\mathbb{E}\{\bs{h}^{(t)}(\bs{q}^{(t)})^{\text{T}}\}=\bs{0}.
\end{align}
This completes the proof of orthogonality of the input and output errors.

\subsection{Relation to Vector AMP}
In this subsection, we will show that OAMP shares the same algorithm as Vector AMP (VAMP) \cite{rangan2019vector}. For convenience, we post the VAMP algorithm by omitting iteration as below
\begin{subequations}
\begin{align}
\hat{\bs{x}}_1&=\left(\sigma_w^{-2}\bs{H}^{\text{T}}\bs{H}+\frac{1}{\gamma_2}\bs{I}\right)^{-1}\left(\sigma_w^{-2}\bs{H}^{\text{T}}\bs{y}+\frac{\bs{r}_2}{\gamma_2}\right),
\label{VAMP-R1}\\
\hat{v}_1&=\frac{1}{N}\text{Tr}\left[\left(\sigma_w^{-2}\bs{H}^{\text{T}}\bs{H}+\frac{1}{\gamma_2}\bs{I}\right)^{-1}\right],
\label{VAMP-R2}\\
\gamma_1&=\left(\frac{1}{\hat{v}_1}-\frac{1}{\gamma_2}\right)^{-1},
\label{VAMP-R3}\\
\bs{r}_1&=\gamma_1\left(\frac{\hat{\bs{x}}_1}{\hat{v}_1}-\frac{\bs{r}_2}{\gamma_2}\right),
\label{VAMP-R4}\\
\hat{\bs{x}}_2&=\mathbb{E}\left\{\bs{x}|\bs{r}_1, \gamma_1\right\},
\label{VAMP-R5}\\
\hat{v}_2&=\frac{1}{N}\sum_{i=1}^N \text{Var}\{x_i|r_{1i}, \gamma_1\},
\label{VAMP-R6}\\
\gamma_2&=\left(\frac{1}{\hat{v}_2}-\frac{1}{\gamma_1}\right)^{-1},
\label{VAMP-R7}\\
\bs{r}_2&=\gamma_2\left(\frac{\hat{\bs{x}}_2}{\hat{v}_2}-\frac{\bs{r}_1}{\gamma_1}\right).
\label{VAMP-R8}
\end{align}
\label{VAMP}
\end{subequations}
\!\!\!Comparing VAMP (\ref{VAMP}) with OAMP in Algorithm \ref{Algorithm:OAMP}, it can be found that equations (\ref{OAMP-R4})-(\ref{OAMP-R7}) of OAMP are equal to equations (\ref{VAMP-R4})-(\ref{VAMP-R8}) of VAMP.
To show the equivalence of OAMP and VAMP, one should prove the equivalence of (\ref{OAMP-R2})-(\ref{OAMP-R3}) and (\ref{VAMP-R3})-(\ref{VAMP-R4}). From (\ref{VAMP-R3}), we have
\begin{align}
\nonumber
\gamma_1&=\frac{\gamma_2\hat{v}_1}{\gamma_2-\hat{v}_1}\\
&=\frac{\gamma_2\text{Tr}\left[\left(\bs{H}^{\text{T}}\bs{H}+\frac{\sigma_w^2}{\gamma_2}\bs{I}\right)^{-1}\right]}{N\frac{\gamma_2}{\sigma_w^2}-\text{Tr}\left[\left(\bs{H}^{\text{T}}\bs{H}+\frac{\sigma_w^2}{\gamma_2}\bs{I}\right)^{-1}\right]},
\label{Equ:G1}
\end{align}
where by SVD $\bs{H}=\bs{U}\bs{\Sigma}\bs{V}^{\text{T}}$
\begin{align}
\nonumber
\text{Tr}\left[\left(\bs{H}^{\text{T}}\bs{H}+\frac{\sigma_w^2}{\gamma_2}\bs{I}\right)^{-1}\right]
&=\text{Tr}\left[\left(\bs{\Sigma}^{\text{T}}\bs{\Sigma}+\frac{\sigma_w^2}{\gamma_2}\right)^{-1}\right]\\
&=\sum_{i=1}^N \frac{\gamma_2}{\lambda_i\gamma_2+\sigma_w^2},
\end{align}
where $\lambda_i$ is the $i$-th eigenvalue of $\bs{H}^{\text{T}}\bs{H}$. Note that if we assume that $\bs{H}^{\text{T}}\bs{H}$ only has $K$ non-zero eigenvalues then $\lambda_i=0$ for $i>K$.

From (\ref{Equ:G1}), we have
\begin{align}
\nonumber
\gamma_1&=\frac{\gamma_2\sum_{i=1}^N \frac{\gamma_2}{\lambda_i\gamma_2+\sigma_w^2}}
{N\frac{\gamma_2}{\sigma_w^2}-\sum_{i=1}^N \frac{\gamma_2}{\lambda_i\gamma_2+\sigma_w^2}}\\
\nonumber
&=\frac{\gamma_2\sum_{i=1}^N \frac{1}{\lambda_i\gamma_2+\sigma_w^2}}{\sum_{i=1}^N \left(\frac{1}{\sigma_w^2}-\frac{1}{\lambda_i\gamma_2+\sigma_w^2}\right)}\\
&=\frac{\gamma_2N-\gamma_2\sum_{i=1}^N \frac{\lambda_i\gamma_2}{\lambda_i\gamma_2+\sigma_w^2}}{\sum_{i=1}^N \frac{\lambda_i\gamma_2}{\lambda_i\gamma_2+\sigma_w^2}}.
\end{align}
On the other hand, from (\ref{OAMP-R2}), we have
\begin{align}
\nonumber
\tau_t^2&=\hat{v}^{(t)}\left(\frac{N}{\sum_{i=1}^N \frac{\lambda_i}{\lambda_i+\frac{\sigma_w^2}{\hat{v}^{(t)}}}}-1\right)\\
&=\frac{N\hat{v}^{(t)}-\hat{v}^{(t)}{\sum_{i=1}^N \frac{\lambda_i\hat{v}^{(t)}}{\hat{v}^{(t)}\lambda_i+\sigma_w^2}}}
{\sum_{i=1}^N \frac{\lambda_i\hat{v}^{(t)}}{\hat{v}^{(t)}\lambda_i+\sigma_w^2}}.
\end{align}
This completes the proof of the equivalence of $\gamma_1$ in VAMP and $\tau_t^2$ in OAMP. In addition, (\ref{Equ:Rt}) and (\ref{Equ:R}) complete the proof of the equivalence of $\bs{r}^{(t)}$ of OAMP and $\bs{r}_1$ of VAMP. Besides, one could find that VAMP algorithm (\ref{VAMP-R1})-(\ref{VAMP-R8}) is same as the diagonal expectation propagation (EP) \cite{minka2001family} and expectation consistent (EC) \cite[Appendix D]{opper2005expectation} (single-loop). They were proposed independently in different manners but shares the same form. Actually, EP/EC (single-loop) with element-wise variance has a slight difference from OAMP/VAMP, where EP/EC (single-loop) is reduced to OAMP/VAMP by taking the mean operation for element-wise variance. The EP was proposed by modifying the assumed density filter, the EC was proposed by minimizing the Gibbs free energy, OAMP was proposed by extending AMP to more general measurement matrix region, and  VAMP was proposed using EP-type message passing. The order of them is EP (2001) by Minka, EC (2005) by Opper, OAMP (2016) by Ma, and VAMP (2016) by Rangan.

\begin{figure*}
\setcounter{equation}{90}
\begin{align}
\nonumber
\bs{r}^{(t)}&=\hat{\bs{x}}^{(t)}+\frac{N}{\text{Tr}(\hat{\bs{W}}_t\bs{H})}\left(\bs{H}^{\text{T}}\bs{H}+\frac{\sigma_w^2}{\hat{v}^{(t)}}\bs{I}\right)^{-1}\bs{H}^{\text{T}}(\bs{y}-\bs{H}\hat{\bs{x}}^{(t)})\\
\nonumber
&=\frac{N}{\text{Tr}(\hat{\bs{W}}_t\bs{H})}\left(\bs{H}^{\text{T}}\bs{H}+\frac{\sigma_w^2}{\hat{v}^{(t)}}\bs{I}\right)^{-1}\bs{H}^{\text{T}}\bs{y}
+\frac{N}{\text{Tr}(\hat{\bs{W}}_t\bs{H})}\left(\bs{H}^{\text{T}}\bs{H}+\frac{\sigma_w^2}{\hat{v}^{(t)}}\bs{I}\right)^{-1}
\left[\frac{\text{Tr}(\hat{\bs{W}}_t\bs{H})}{N}\left(\bs{H}^{\text{T}}\bs{H}+\frac{\sigma_w^2}{\hat{v}^{(t)}}\bs{I}\right)\hat{\bs{x}}^{(t)}-\bs{H}^{\text{T}}\bs{H}\hat{\bs{x}}^{(t)}\right]\\
\nonumber
&=\frac{\left(\bs{H}^{\text{T}}\bs{H}+\frac{\sigma_w^2}{\hat{v}^{(t)}}\bs{I}\right)^{-1}\bs{H}^{\text{T}}\bs{y}}
{\frac{1}{N}\sum_{i=1}^N \frac{\lambda_i\hat{v}^{(t)}}{\hat{v}^{(t)}\lambda_i+\sigma_w^2}}
+\frac{\left(\bs{H}^{\text{T}}\bs{H}+\frac{\sigma_w^2}{\hat{v}^{(t)}}\bs{I}\right)^{-1}}{\frac{1}{N}\sum_{i=1}^N \frac{\lambda_i\hat{v}^{(t)}}{\hat{v}^{(t)}\lambda_i+\sigma_w^2}}
\left[\left(\frac{1}{N}\sum_{i=1}^N \frac{\lambda_i\hat{v}^{(t)}}{\hat{v}^{(t)}\lambda_i+\sigma_w^2}-1\right)\bs{H}^{\text{T}}\bs{H}\hat{\bs{x}}^{(t)}+
\frac{1}{N}\sum_{i=1}^N \frac{\lambda_i\sigma_w^2}{\hat{v}^{(t)}\lambda_i+\sigma_w^2}\hat{\bs{x}}^{(t)}
\right]\\
&=\frac{\left(\bs{H}^{\text{T}}\bs{H}+\frac{\sigma_w^2}{\hat{v}^{(t)}}\bs{I}\right)^{-1}\bs{H}^{\text{T}}\bs{y}}
{\frac{1}{N}\sum_{i=1}^N \frac{\lambda_i\hat{v}^{(t)}}{\hat{v}^{(t)}\lambda_i+\sigma_w^2}}+
\left(\bs{H}^{\text{T}}\bs{H}+\frac{\sigma_w^2}{\hat{v}^{(t)}}\bs{I}\right)^{-1}
\left(\frac{\sigma_w^2}{\hat{v}^{(t)}}\hat{\bs{x}}^{(t)}-\frac{\sigma_w^2\frac{1}{N}\sum_{i=1}^N \frac{1}{\hat{v}^{(t)}\lambda_i+\sigma_w^2}}{\hat{v}^{(t)}\frac{1}{N}\sum_{i=1}^N \frac{\lambda_i}{\hat{v}^{(t)}\lambda_i+\sigma_w^2}}\bs{H}^{\text{T}}\bs{H}\hat{\bs{x}}^{(t)}\right).
\label{Equ:Rt}
\end{align}
\hrulefill
\begin{align}
\nonumber
\bs{r}_1&=\frac{\gamma_2\hat{\bs{x}}_1}{\gamma_2-\hat{v}_1}-\frac{\hat{v}_1\bs{r}_2}{\gamma_2-\hat{v}_1}\\
\nonumber
&=\frac{\gamma_2\left(\sigma_w^{-2}\bs{H}^{\text{T}}\bs{H}+\frac{1}{\gamma_2}\bs{I}\right)^{-1}\left(\sigma_w^{-2}\bs{H}^{\text{T}}\bs{y}+\frac{\bs{r}_2}{\gamma_2}\right)}
{\gamma_2-\frac{1}{N}\text{Tr}\left[\left(\sigma_w^{-2}\bs{H}^{\text{T}}\bs{H}+\frac{1}{\gamma_2}\bs{I}\right)^{-1}\right]}
-\frac{\frac{1}{N}\text{Tr}\left[\left(\sigma_w^{-2}\bs{H}^{\text{T}}\bs{H}+\frac{1}{\gamma_2}\bs{I}\right)^{-1}\right]\bs{r}_2}{\gamma_2-\frac{1}{N}\text{Tr}\left[\left(\sigma_w^{-2}\bs{H}^{\text{T}}\bs{H}+\frac{1}{\gamma_2}\bs{I}\right)^{-1}\right]}\\
\nonumber
&=\frac{\gamma_2\left(\bs{H}^{\text{T}}\bs{H}+\frac{\sigma_w^2}{\gamma_2}\bs{I}\right)^{-1}\left(\bs{H}^{\text{T}}\bs{y}+\frac{\sigma_w^2}{\gamma_2}\bs{r}_2\right)}
{\gamma_2-\frac{\sigma_w^2}{N}\sum_{i=1}^N \frac{\gamma_2}{\lambda_i\gamma_2+\sigma_w^2}}
-\frac{\frac{\sigma_w^2}{N}\sum_{i=1}^N \frac{\gamma_2}{\lambda_i\gamma_2+\sigma_w^2}\bs{r}_2}{\gamma_2-\frac{\sigma_w^2}{N}\sum_{i=1}^N \frac{\gamma_2}{\lambda_i\gamma_2+\sigma_w^2}}\\
\nonumber
&=\frac{\left(\bs{H}^{\text{T}}\bs{H}+\frac{\sigma_w^2}{\gamma_2}\bs{I}\right)^{-1}\bs{H}^{\text{T}}\bs{y}}{\frac{1}{N}\sum_{i=1}^N \frac{\lambda_i\gamma_2}{\lambda_i\gamma_2+\sigma_w^2}}
+\left(\bs{H}^{\text{T}}\bs{H}+\frac{\sigma_w^2}{\gamma_2}\bs{I}\right)^{-1}
\left[\frac{\frac{\sigma_w^2}{\gamma_2}}{\frac{1}{N}\sum_{i=1}^N \frac{\lambda_i\gamma_2}{\lambda_i\gamma_2+\sigma_w^2}}\bs{r}_2-
\left(\bs{H}^{\text{T}}\bs{H}+\frac{\sigma_w^2}{\gamma_2}\bs{I}\right)\frac{\frac{1}{N}\sum_{i=1}^N \frac{\sigma_w^2}{\lambda_i\gamma_2+\sigma_w^2}}{\frac{1}{N}\sum_{i=1}^N \frac{\lambda_i\gamma_2}{\gamma_2\lambda_i+\sigma_w^2}}
\bs{r}_2\right]\\
&=\frac{\left(\bs{H}^{\text{T}}\bs{H}+\frac{\sigma_w^2}{\gamma_2}\bs{I}\right)^{-1}\bs{H}^{\text{T}}\bs{y}}{\frac{1}{N}\sum_{i=1}^N \frac{\lambda_i\gamma_2}{\lambda_i\gamma_2+\sigma_w^2}}
+\left(\bs{H}^{\text{T}}\bs{H}+\frac{\sigma_w^2}{\gamma_2}\bs{I}\right)^{-1}\left(\frac{\sigma_w^2}{\gamma_2}\bs{r}_2
-\frac{\sigma_w^2\frac{1}{N}\sum_{i=1}^N \frac{1}{\lambda_i\gamma_2+\sigma_w^2}}{\gamma_2\frac{1}{N}\sum_{i=1}^N \frac{\lambda_i}{\gamma_2\lambda_i+\sigma_w^2}}\bs{H}^{\text{T}}\bs{H}\bs{r}_2\right).
\label{Equ:R}
\end{align}
\hrulefill
\end{figure*}

\subsection{State Evolution}
The asymptotic MSE of OAMP is defined as
\begin{align}
\nonumber
\textsf{mse}(\bs{x},t+1)&=\lim_{N\rightarrow \infty}\frac{1}{N}\|\hat{\bs{x}}_{\text{mmse}}^{(t+1)}-\bs{x}\|_2^2\\
\nonumber
&=\lim_{N\rightarrow \infty}\frac{1}{N}\sum_{i=1}^N (\hat{x}_{\text{mmse},i}^{(t+1)}-x_i)^2\\
&\overset{\text{a.s.}}{=}\mathbb{E}_{\textsf{X},\textsf{Z}}\left\{\left(\eta_t^{\text{mmse}}(\textsf{X}+\tau_t\textsf{Z})-\textsf{X}\right)^2\right\}.
\label{Equ:L1}
\end{align}
where the last equation holds by Assumption 1. As we observed from OAMP in Algorithm~\ref{Algorithm:OAMP}, in the large system limit, the variance of OAMP estimator can be written as
\begin{align}
\nonumber
\hat{v}^{(t+1)}_{\text{mmse}}&=\frac{1}{N}\sum_{i=1}^N\text{Var}\{x_i|r_i^{(t)}, \tau_t\}\\
&\overset{\text{a.s.}}{=}\mathbb{E}_{\textsf{X},\textsf{Z}}\left\{\left(\eta_t^{\text{mmse}}(\textsf{X}+\tau_t\textsf{Z})-\textsf{X}\right)^2\right\}.
\label{Equ;L2}
\end{align}
Combining (\ref{Equ:L1}) and (\ref{Equ;L2}) proves that the variance of OAMP estimator is equal to asymptotic MSE of OAMP almost sure, i.e., $\hat{v}^{(t+1)}_{\text{mmse}}\overset{\text{a.s.}}{=}\textsf{mse}(\bs{x},t+1)$. Note that $\hat{v}_{\text{mmse}}^{(t+1)}$ in (\ref{Equ;L2}) only relies on the parameter $\tau_t^2$ and this parameter can be obtained by
\begin{align}
\hat{v}^{(t)}&=\left(\frac{1}{\hat{v}^{(t)}_{\text{mmse}}}-\frac{1}{\tau_{t-1}^2}\right)^{-1},\\
\tau_t^2&=\hat{v}^{(t)}\left(\frac{N}{\text{Tr}(\hat{\bs{W}}_t\bs{H})}-1\right),
\end{align}
where by SVD $\bs{H}=\bs{U\Sigma V}^{\text{T}}$
\begin{align}
\nonumber
\frac{1}{N}\text{Tr}(\hat{\bs{W}}_t\bs{H})
&=\frac{1}{N}\text{Tr}\left(\bs{H}^{\text{T}}\left(\bs{HH}^{\text{T}}+\frac{\sigma^2}{\hat{v}^{(t)}}\bs{I}\right)^{-1}\bs{H}\right)\\
\nonumber
&=\frac{1}{N}\text{Tr}\left(\bs{\Sigma}^{\text{T}}\left(\bs{\Sigma\Sigma}^{\text{T}}+\frac{\sigma^2}{\hat{v}^{(t)}}\bs{I}\right)^{-1}\bs{\Sigma}\right)\\
\nonumber
&=\frac{1}{N}\sum_{i=1}^N \frac{\sigma_i^2}{\sigma_i^2+\frac{\sigma_w^2}{\hat{v}^{(t)}}}\\
&\overset{\text{a.s.}}{=}\mathbb{E}\left\{\frac{\lambda}{\lambda+\frac{\sigma_w^2}{\hat{v}^{(t)}}}\right\},
\end{align}
where $\sigma_i$ is the $i$-th diagonal element of $\bs{\Sigma}$, and the expectation in $\mathbb{E}\{\lambda\}$ is taken over the asymptotic eigenvalue distribution of $\bs{H}^{\text{T}}\bs{H}$.

In the sequel, we obtain the SE of OAMP as below
\begin{align}
&\text{LE:\ } \ \quad \ \tau_t^2=\hat{v}^{(t)}\left(\mathbb{E}\left\{\frac{\lambda^2}{\lambda^2+\frac{\sigma_w^2}{\hat{v}^{(t)}}}\right\}^{-1}-1\right)\\
&\text{NLE:}\quad
\begin{cases}
\hat{v}^{(t+1)}_{\text{mmse}}=\mathbb{E}_{\textsf{X},\textsf{Z}}\left\{\left(\eta_t^{\text{mmse}}(\textsf{X}+\tau_t\textsf{Z})-\textsf{X}\right)^2\right\}\\
\hat{v}^{(t+1)}=\left(\frac{1}{\hat{v}^{(t+1)}_{\text{mmse}}}-\frac{1}{\tau_{t}^2}\right)^{-1}
\end{cases}
\end{align}
Be aware, in the NLE part, the $\hat{v}_{\text{mmse}}^{(t+1)}$ is output MSE rather than $\hat{v}^{(t+1)}$.

\subsection{Numeric Simulations}
\begin{figure}[!t]
\centering
\includegraphics[width=0.43\textwidth]{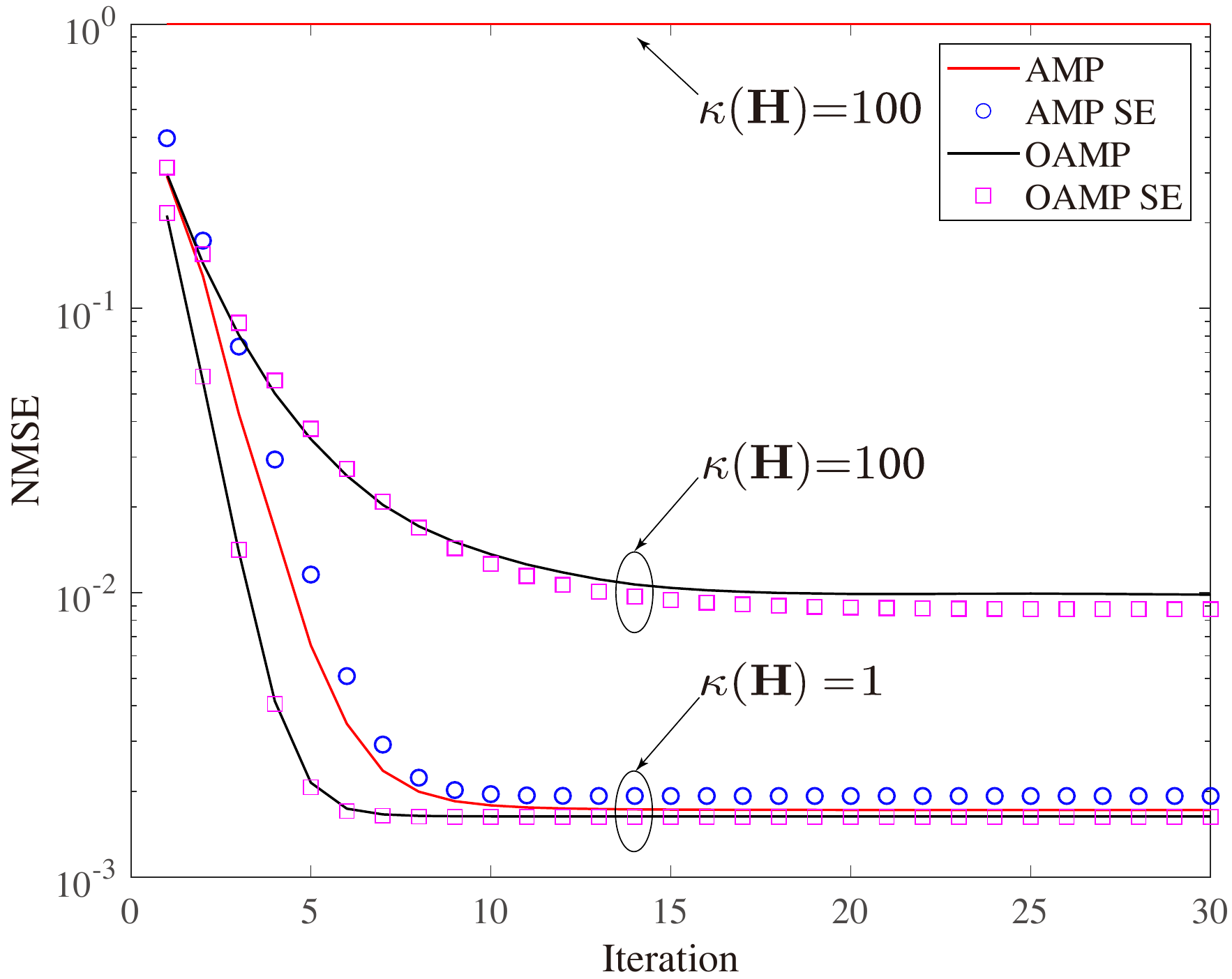}
\caption{Iterative behavior of OAMP, AMP and their SEs in compressed sensing. $M=512$, $N=1024$ and SNR=$1/\sigma_w^2$=20dB. $\bs{x}$ has IID BG entries following $\mathcal{BG}(0, 0.05)$. The measurement matrix is generated by $\bs{H}=\bs{U}\bs{\Sigma}\bs{V}^{\text{T}}$  where $\bs{U}$ and $\bs{V}$ are both Haar distribution and $\bs{\Sigma}$ is rectangular matrix whose diagonal is $\sigma_1\cdots \sigma_M$ with $\sigma_i/\sigma_{i+1}=\kappa(\bs{H})^{1/M}$ and $\sum_{i=1}^M \sigma_i^2=N$ such that $|h_{ai}|^2=O(1/M)$. The condition number is defined as $\kappa(\bs{H})=\frac{\sigma_{\text{max}}(\bs{H})}{\sigma_{\text{min}}(\bs{H})}$ where $\sigma_{\text{max}}(\bs{H})$ and $\sigma_{\min}(\bs{H})$ denotes maximum and minimum  singular values of $\bs{H}$, respectively.
}
\label{EXP3SE}
\end{figure}

In Fig~\ref{EXP3SE}, we present the comparison of OAMP, AMP and their SEs in compressed sensing. In $\kappa(\bs{H})=1$, the SE curves match AMP or OAMP well and OAMP converges faster than AMP. In this case, the gap of the fixed point of AMP and OAMP can be ignored. On the other hand, in $\kappa(\bs{H})=100$, AMP fails to converge while OAMP and its SE converge to the same fixed point. Besides, the MSE performance of OAMP in $\kappa(\bs{H})=1$ is better than that in $\kappa(\bs{H})=100$ (ill-conditioned matrix). Note that since the SE of AMP is obtained under the Gaussian random matrix and thus the condition number has no effect on the performance of SE of AMP.

\section{Long Memory AMP}
Although OAMP can be applied to more general random matrices, its complexity with roughly $\mathcal{O}(N^3)$ is larger than AMP with roughly $\mathcal{O}(N^2)$. To balance the computational complexity and region of random measurement matrix, several long memory algorithms have been proposed, such as convolution AMP (CAMP) \cite{takeuchi2021bayes} and memory AMP (MAMP) \cite{liu2021memory}. CAMP only adjusts the Onsager term where all preceding messages are involved to ensure the Gaussianity of input error. However, CAMP may fail to convergence in ill-conditioned measurement matrix such as large conditional number. Following CAMP and OAMP, MAMP applies a few terms of matrix Taylor series to carry out the matrix inversion in OAMP and modifies the structure of input signal of denoiser to ensure  (a) the orthogonality of all preceding input errors and $t$-th output error, (b) the orthogonality of $t$-th input error and original signal $\bs{x}$, (c) the orthogonality of $t$-th input error and all preceding output errors.

Recalling the OAMP iterations in Algorithm~\ref{Algorithm:OAMP}, the complexity of OAMP is dominated by the matrix inversion in (\ref{OAMP-R1}). Let's define $\varsigma_t=\frac{\sigma_w^2}{\hat{v}^{(t)}}$ and a relaxation parameter $\theta_t$. Then
\begin{align}
\left(\theta_t\left(\bs{H}\bs{H}^{\text{T}}+\varsigma_t\bs{I}\right)\right)^{-1}
&=\left(\bs{I}-\left(\bs{I}-\theta_t(\bs{H}\bs{H}^{\text{T}}+\varsigma_t\bs{I})\right)\right)^{-1}.
\end{align}
Defining $\bs{C}_t=\bs{I}-\theta_t(\bs{H}\bs{H}^{\text{T}}+\varsigma_t\bs{I})$, we have
\begin{align}
\left(\theta_t\left(\bs{H}\bs{H}^{\text{T}}+\varsigma_t\bs{I}\right)\right)^{-1}
=(\bs{I}-\bs{C}_t)^{-1}.
\end{align}
As spectral radius of $\bs{C}_t$ satisfies $\rho(\bs{C}_t)<1$,  applying matrix Taylor series gets
\begin{align}
\left(\theta_t\left(\bs{H}\bs{H}^{\text{T}}+\varsigma_t\bs{I}\right)\right)^{-1}
&=\sum_{k=0}^{\infty}\bs{C}_t^k.
\end{align}
It can be verified that $\theta_t=(\lambda^{\dag}+\varsigma_t)^{-1}$ with $\lambda^{\dag}=\frac{\lambda_{\text{max}}+\lambda_{\text{min}}}{2}$ satisfies $\rho(\bs{C}_t)<1$, where $\lambda_{\text{max}}$ and $\lambda_{\text{min}}$ denote the maximum and minimum eigenvalue of $\bs{H}\bs{H}^{\text{T}}$, respectively. For convenience, defining $\bs{B}=\lambda^{\dag}\bs{I}-\bs{HH}^{\text{T}}$ yields
\begin{align}
\hat{\bs{W}}_t(\bs{y}-\bs{H}\hat{\bs{x}}^{(t)})
=\bs{H}^{\text{T}}\sum_{k=0}^{\infty}(\theta_t\bs{B})^k(\bs{y}-\bs{H}\hat{\bs{x}}^{(t)}).
\end{align}
However, the complexity of the exact approximation is still huge. The MAMP applies a few terms of matrix series to represent matrix inversion and use all preceding terms to ensure three orthogonality.

The MAMP considers the following structure:
\begin{subequations}
\begin{align}
&\text{LE}: \quad \
\begin{cases}
\bs{z}^{(t)}=\theta_t\bs{B}\bs{z}^{(t-1)}+\xi_t(\bs{y}-\bs{H}\hat{\bs{x}}^{(t)})\\
\bs{r}^{(t)}=\frac{1}{\varepsilon_t}\left(\bs{H}^{\text{T}}\bs{z}^{(t)}+\sum_{i=1}^tp_{t,i}\hat{\bs{x}}^{(i)}\right)
\end{cases},
\label{MAMP_A}\\
&\text{NLE}: \
\begin{cases}
\hat{\bs{x}}_{\text{mmse}}^{(t+1)}&=\eta_t^{\text{mmse}}(\bs{r}^{(t)}, \tau_{t,t})\\
\hat{v}^{(t+1)}_{\text{mmse}}&=\frac{1}{N}\sum_{i=1}^N\text{Var}\{x_i|r_i^{(t)}, \tau_{t,t}\}\\
\hat{v}_{t+1,t+1}&=\left(\frac{1}{\hat{v}^{(t+1)}_{\text{mmse}}}-\frac{1}{\tau_{t,t}^2}\right)^{-1}\\
\hat{\bs{x}}^{(t+1)}&=\hat{v}_{t+1,t+1}\left(\frac{\hat{\bs{x}}_{\text{mmse}}^{(t+1)}}{\hat{v}^{(t+1)}_{\text{mmse}}}-\frac{\bs{r}^{(t)}}{\tau_{t,t}^2}\right)^{-1}
\end{cases},
\label{MAMP_B}
\end{align}
\end{subequations}
where $\bs{B}=\lambda^{\dag}\bs{I}-\bs{HH}^{\text{T}}$ and $\theta_t=(\lambda^{\dag}+\varsigma_t)^{-1}$. Note that $\hat{\bs{x}}_{\text{mmse}}^{(t+1)}$ is the output estimator rather than $\hat{\bs{x}}^{(t+1)}$.

\begin{remark}
As can be seen from the MAMP algorithm in (\ref{MAMP_A})-(\ref{MAMP_B}), the parameter $\lambda^{\dag}=\frac{\lambda_{\text{min}}+\lambda_{\text{max}}}{2}$ of  MAMP algorithm relies on the eigenvalue of $\bs{HH}^{\text{T}}$ which is roughly with cost of $\mathcal{O}(N^3)$. Although some works give the approximations to the maximum or minimum singular value of $\bs{H}$, its complexity is still huge.  We also note that in the long version \cite{liu2020memory_long}, a simple bound of maximum eigenvalue and minimum eigenvalue is applied to provide a close performance of perfect eigenvalues, especially in low condition number. In the case of given eigenvalues of $\bs{HH}^{\text{T}}$, the MAMP balances the computational complexity and random measurement region well.
\end{remark}

\subsection{Derivation of MAMP}

Similar to OAMP, the following assumptions are applied
\begin{itemize}
\item \textbf{Assumption 3}: the input error $\bs{h}^{(t)}$ consists of IID zero-mean Gaussian entries independent of $\bs{x}$, i.e., $\textsf{R}^{(t)}=\textsf{X}+\tau_{t,t}\textsf{Z}_t$ with $\textsf{Z}$ being a standard Gaussian RV. Let's define $\eta_t=\tau_{t,t}\textsf{Z}_t$. Different from OAMP, MAMP assumes that  $[\eta_1, \cdots, \eta_t]^{\text{T}}$ follows joint Gaussian $\mathcal{N}(\bs{0}, \bs{V}_t)$ with $\bs{V}_t=[\tau_{i,j}^2]_{t\times t}$.
\item \textbf{Assumption 4}: the output error $\bs{q}^{(t+1)}$ consists of IID entries independent of $\bs{H}$ and noise $\bs{n}$.
\end{itemize}

Using initial conditions $\bs{z}^{(0)}=\hat{\bs{x}}^{(0)}=\bs{0}$, from (\ref{MAMP_A})
\begin{align}
\bs{z}^{(t)}&=\sum_{i=1}^t \xi_i \left(\prod_{j=i+1}^t\theta_j\right)\bs{B}^{t-i}(\bs{y}-\bs{H}\hat{\bs{x}}^{(i)}).
\end{align}
Defining $\overline{\theta}_{t,i}=\prod_{j=i+1}^t\theta_j$ ($\overline{\theta}_{t,i}=1$ for $i\geq t$), we have
\begin{align}
\bs{r}^{(t)}&=\frac{1}{\varepsilon_t}\left(
\bs{Q}_t\bs{y}+\sum_{i=1}^t\bs{H}_i^t\hat{\bs{x}}^{(i)}
\right),
\end{align}
where
\begin{align}
\bs{Q}_t&=\sum_{i=1}^t\xi_i\overline{\theta}_{t,i}\bs{H}^{\text{T}}\bs{B}^{t-i},\\
\bs{H}_i^t&=p_{t,i}\bs{I}-\xi_i\overline{\theta}_{t,i}\bs{H}^{\text{T}}\bs{B}^{t-i}\bs{H}.
\end{align}

From the orthogonality of input error and original signal i.e., $\frac{1}{N}(\bs{h}^{(t)})^{\text{T}}\bs{x}\overset{\text{a.s.}}{=}0$, we have
\begin{align}
\nonumber
&\frac{1}{N}\left(\bs{h}^{(t)}\right)^{\text{T}}\bs{x}\\
\nonumber
&=\frac{1}{N}\left(\frac{1}{\varepsilon_t}\bs{Q}_t(\bs{Hx}+\bs{n})+\frac{1}{\varepsilon_t}\sum_{i=1}^t\bs{H}_i^t(\bs{q}^{(i)}+\bs{x})-\bs{x}\right)^{\text{T}}\bs{x}\\
&=\frac{1}{N\varepsilon_t}\bs{x}^{\text{T}}((\bs{Q}_t\bs{H})^{\text{T}}+\sum_{i=1}^{t}(\bs{H}_i^t)^{\text{T}})\bs{x}-\frac{1}{N}\bs{x}^{\text{T}}\bs{x},
\end{align}
where $\frac{1}{N}(\bs{q}^{(i)})^{\text{T}}\bs{x}\overset{\text{a.s.}}{=}0$ is applied. Then we get
\begin{align}
\frac{1}{N\varepsilon_t}\text{Tr}\left\{\bs{Q}_t\bs{H}+\sum_{i=1}^{t}\bs{H}_i^t\right\}=1.
\label{Equ:J1}
\end{align}
From the orthogonality of input error and output errors, i.e., $\frac{1}{N}(\bs{h}^{(t)})^{\text{T}}\bs{q}^{(i)}\overset{\text{a.s.}}{=}0$, we have
\begin{align}
\text{Tr}\{\bs{H}_i^t\}=0.
\label{Equ:J2}
\end{align}

Combining (\ref{Equ:J1}) and (\ref{Equ:J2}), we have
\begin{align}
p_{t,i}&=\frac{1}{N}\xi_i \overline{\theta}_{t,i}\text{Tr}\left\{\bs{H}^{\text{T}}\bs{B}^{t-i}\bs{H}\right\}\\
\varepsilon_{t}&=\sum_{i=1}^tp_{t,i}
\end{align}
where the parameters $p_{t,i}$ and $\varepsilon_t$ can be determined once the parameters $\{\theta_t\}$ and $\{\xi_t\}$ are determined, where $\xi_t$ is obtained by minimizing the averaged input error
\begin{align}
\tau_{t,t}^2=\lim_{N\rightarrow \infty}\frac{1}{N}\|\bs{r}^{(t)}-\bs{x}\|_2^2.
\end{align}
Using the facts $\frac{1}{N}(\bs{q}^{(i)})^{\text{T}}\bs{x}$ and independence of $\bs{n}$ and $\bs{x}$, we have
\begin{align}
\nonumber
\tau_{t,t}^2&=\frac{1}{N\varepsilon_t^2}\left\|\bs{Q}_t\bs{n}+\sum_{i=1}^N\bs{H}_i^t\bs{q}^{(i)}\right\|_2^2\\
\nonumber
&=\frac{1}{N\varepsilon_t^2}\left(\sum_{i=1}^t\sum_{j=1}^t\xi_i\xi_j\theta_{t,i}\theta_{t,j}\sigma_w^2\text{Tr}\left\{\bs{H}^{\text{T}}\bs{B}^{2t-i-j}\bs{H}\right\}
\right.\\
&\qquad \qquad -\left.\sum_{i=1}^t\sum_{j=1}^t\hat{v}_{i,j}\text{Tr}\left\{(\bs{H}_i^t)^{\text{T}}\bs{H}_i^t\right\}\right),
\end{align}
where $\hat{v}_{i,j}=\frac{1}{N}(\bs{q}^{(i)})^{\text{T}}\bs{q}^{(j)}$ and $\hat{v}_{i,j}=\hat{v}_{j,i}$. Defining
\begin{align}
\vartheta_{t,i}&=\xi_i\overline{\theta}_{t,i},
\label{Equ:K1}\\
\bs{W}_t&=\bs{H}^{\text{T}}\bs{B}^t\bs{H}, \  \ \ w_t=\frac{1}{N}\text{Tr}(\bs{W}_t),\\
\bs{N}_{i,j}&=\bs{W}_{i}\bs{W}_j, \ \ \overline{w}_{i,j}=\frac{1}{N}\text{Tr}\left\{\bs{N}_{i,j}\right\}-w_{i}w_{j},
\end{align}
we get $p_{t,i}=\vartheta_{t,i}w_{t-i}$ and
\begin{align}
\nonumber
\tau_{t,t}^2
&=\frac{1}{\varepsilon_t^2}\sum_{i=1}^t\sum_{j=1}^t\vartheta_{t,i}\vartheta_{t,j}\left(\sigma_w^2w_{2t-i-j}
+\hat{v}_{i,j}\overline{w}_{t-i,t-j}\right)\\
&=\frac{c_{t,1}\xi_t^2-2c_{t,2}\xi_t+c_{t,3}}{w_0^2(\xi_t+c_{t,0})^2},
\label{Equ:L2}
\end{align}
where
\begin{align*}
c_{t,0}&=\sum_{i=1}^{t-1}\frac{p_{t,i}}{w_0},\\
c_{t,1}&=\sigma_w^2w_0+\hat{v}_{t,t}\overline{w}_{0,0},\\
c_{t,2}&=-\sum_{i=1}^{t-1}\vartheta_{t,i}(\sigma_w^2w_{t-i}+\hat{v}_{t,i}\overline{w}_{0,t-i}),\\
c_{t,3}&=\sum_{i=1}^{t-1}\sum_{j=1}^{t-1}\vartheta_{t,i}\vartheta_{t,j}(\sigma_w^2w_{2t-i-j}+\hat{v}_{i,j}\overline{w}_{t-i,t-j}).
\end{align*}
The parameter $\xi_t$ is obtained by minimizing $\tau_{t,t}^2$. Zeroing $\frac{\partial \tau_{t,t}^2}{\partial \xi_t}$ gets two points $\xi_t=-c_{t,0}$ and
$
\xi_t=\frac{c_{t,2}c_{t,0}+c_{t,3}}{c_{t,1}c_{t,0}+c_{t,2}}
$,
where $\xi_t=-c_{t,0}$ is maximum value point while
\begin{align}
\xi_t^{\star}=\begin{cases}
\frac{c_{t,2}c_{t,0}+c_{t,3}}{c_{t,1}c_{t,0}+c_{t,2}}  &c_{t,1}c_{t,0}+c_{t,2}\ne 0\\
+\infty   &\text{otherwise}
\end{cases}.
\end{align}

Defining the residual error $
\tilde{\bs{z}}^{(t)}=\bs{y}-\bs{H}\hat{\bs{x}}^{(t)}
$,
the crossed variance $\hat{v}_{i,j}$ can be provided by
\begin{align}
\nonumber
\frac{1}{N}(\tilde{\bs{z}}^{(i)})^{\text{T}}\tilde{\bs{z}}^{(j)}
&=\frac{1}{N}\left[\bs{H}(\bs{x}-\hat{\bs{x}}^{(i)})+\bs{n}\right]^{\text{T}}\left[\bs{H}(\bs{x}-\hat{\bs{x}}^{(j)})+\bs{n}\right]\\
\nonumber
&=\frac{1}{N}\left(-\bs{H}\bs{q}^{(i)}+\bs{n}\right)^{\text{T}}\left(-\bs{H}\bs{q}^{(j)}+\bs{n}\right)\\
&=\frac{1}{N}\text{Tr}\left\{\bs{H}^{\text{T}}\bs{H}\right\}\hat{v}_{i,j}+\alpha \sigma_w^2.
\end{align}
It implies $\hat{v}_{i,j}=(\frac{1}{N}(\tilde{\bs{z}}^{(i)})^{\text{T}}\tilde{\bs{z}}^{(j)}-\alpha\sigma_w^2)/w_0$.

From (\ref{Equ:K1}), we get
\begin{align}
\vartheta_{t,i}=
\begin{cases}
\theta_t\vartheta_{t-1,i} & 0\leq i<t-1\\
\xi_{t-1}\theta_t &i=t-1\\
\xi_t    & i=t
\end{cases}.
\end{align}
Totally, the MAMP is run in the following steps: (a) calculating parameters: $\theta_t$, $\vartheta_{t,i} $, $p_{t,i}$  $(i<t)$; (b) calculating parameters: $c_{t,0}$, $c_{t,1}$, $c_{t,2}$, and $c_{t,3}$, and applying them to get $\xi_t=\vartheta_{t,t}$, $p_{t,t}$, and $\varepsilon_t$; (c) calculating $\tau_{t,t}^2$ and carrying out LE; (d) carrying out NLE and calculating $[\hat{v}_{i,j}]_{(t+1)\times (t+1)}$.

In fact, the MAMP is easy to fail to converge without damping, especially in the case of large condition number (e.g., $\kappa(\bs{H})>10^2$). To ensure the convergence of MAMP, the damping factor is applied to the parameters $\hat{\bs{x}}^{(t+1)}$, $\hat{v}_{t+1,t+1}$, and  $\tilde{\bs{z}}^{(t+1)}$
\begin{align*}
\hat{\bs{x}}^{(t+1)}&=\beta_1^{(t)}\hat{\bs{x}}^{(t+1)}+(1-\beta_1^{(t)})\hat{\bs{x}}^{(t)},\\
\tilde{\bs{z}}^{(t+1)}&=\beta_1^{(t)}\tilde{\bs{z}}^{(t+1)}+(1-\beta_1^{(t)})\tilde{\bs{z}}^{(t+1)},\\
\hat{v}_{t+1,i}&=\beta_2^{(t)}\hat{v}_{t+1,i}+(1-\beta_2^{(t)})\hat{v}_{t+1,i-1},
\end{align*}
for $1< i<t+1$. Different from the damping presented here, \cite{liu2021memory} shows another kind damping. But, in fact, the damping factor only has the effect on the convergence speed if algorithm converges.

\subsection{State Evolution}
Similar to other AMP-like algorithms, the MSE of MAMP can also be predicted by its SE. The asymptotic MSE of MAMP is defined as
\begin{align}
\nonumber
\textsf{mse}(\bs{x},t+1)
&=\frac{1}{N}\|\hat{\bs{x}}_{\text{mmse}}^{(t+1)}-\bs{x}\|_2^2\\
\nonumber
&\overset{\text{a.s.}}{=}\mathbb{E}\left\{(\eta_t^{\text{mmse}}(\textsf{X}+\tau_{t,t}\textsf{Z})-\textsf{X})^2\right\}\\
&\overset{\text{a.s.}}{=}\hat{v}_{t+1,t+1}.
\end{align}
This term only relies on the parameter $\tau_{t,t}^2$, which can be obtained by (\ref{Equ:L2}). In $\tau_{t,t}^2$, the parameter $\hat{v}_{i,j}\overset{\text{a.s.}}{=}\frac{1}{N}(\bs{q}^{(i)})^{\text{T}}\bs{q}^{(j)}$ is obtained numerically by generating $x$ following $\pa_{\textsf{X}}(x)=\rho\mathcal{N}(x|0,\rho^{-1})+(1-\rho)\delta(x)$ and $r^{(t)}=x+n_t$ with $[n_1,\cdots, n_t]\sim \mathcal{N}(\bs{0}, \bs{\Xi}_t)$ where $\bs{\Xi}_t=[\tau_{i,j}^2]_{t\times t}$ and
\begin{align*}
\nonumber
\tau_{t,\tau}^2
&=\lim_{N\rightarrow \infty}\frac{1}{N}(\bs{r}^{(t)}-\bs{x})^{\text{T}}(\bs{r}^{(\tau)}-\bs{x})\\
&=\frac{1}{\varepsilon_t\varepsilon_\tau}\sum_{i=1}^t\sum_{j=1}^{\tau}\vartheta_{t,i}\vartheta_{\tau,j}\left(\sigma_w^2w_{t+\tau-i-j}
+\hat{v}_{ij}\overline{w}_{t-i,\tau-j}\right),
\end{align*}
with $\tau_{t,\tau}=\tau_{\tau,t}$. Then,
\begin{align*}
\forall \tau<t:\quad \hat{v}_{t,\tau}=\mathbb{E}\left\{(\hat{x}^{(t)}-x)(\hat{x}^{(\tau)}-x)\right\}.
\end{align*}

\subsection{Numeric Simulation}
\begin{figure}[!t]
\centering
\includegraphics[width=0.43\textwidth]{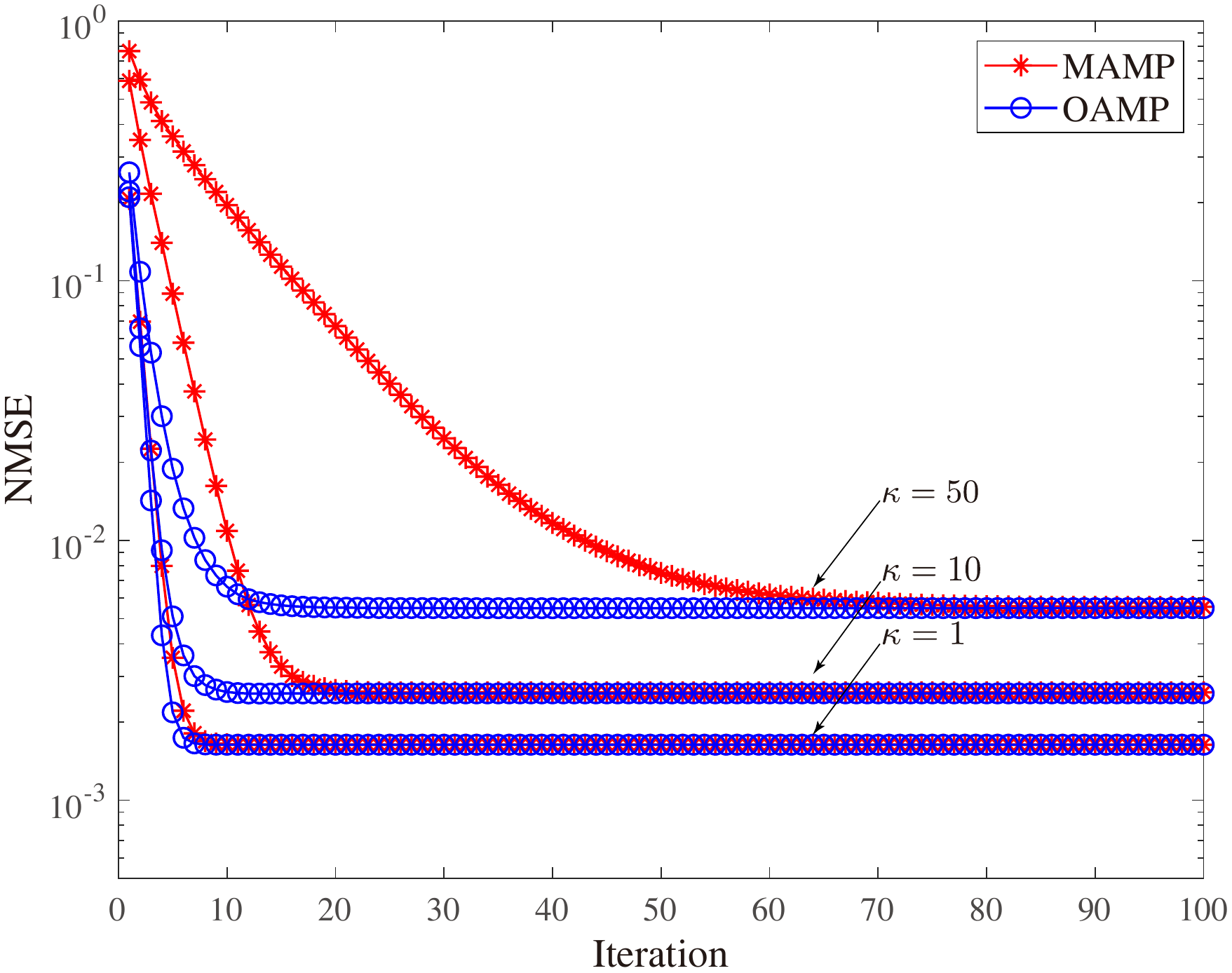}
\caption{Comparison of MAMP and OAMP in different condition numbers. Each entry of $\bs{x}$ is generated from BG distribution $\mathcal{BG}(0,0.1)$. $(M,N)=(1024, 512)$ and $\text{SNR}=1/\sigma_w^2=20\text{dB}$. The measurement matrix is generated by $\bs{H}=\bs{U\Sigma V}^{\text{T}}$ where both $\bs{U}$ and $\bs{V}$ are Haar-distributed and $\bs{\Sigma}$ is rectangular matrix whose diagonal elements are $\sigma_1,\cdots, \sigma_M$ with $\frac{\sigma_i}{\sigma_{i+1}}=\kappa(\bs{H})^{1/M}$ and $\sum_{i=1}^M\sigma_i^2=N$, where $\kappa(\bs{H})=\frac{\sigma_{\text{max}}(\bs{H})}{\sigma_{\text{min}}(\bs{H})}$ with $\sigma_{\text{max}}(\bs{H})$ and $\sigma_{\text{min}}(\bs{H})$ being maximum and minimum singular values of $\bs{H}$, respectively. The damping factors $\beta_1^{(t)}=0.7$ and $\beta_2^{(t)}=0.8$ are applied to the cases of $\kappa(\bs{H})=1$ and $\kappa(\bs{H})=10$, while $\beta_1^{(t)}=\beta_2^{(t)}=0.4$ are applied to the case of $\kappa(\bs{H})=50$.
}
\label{EXP4}
\end{figure}

In Fig.~\ref{EXP4}, we show the pre-iteration behavior of MAMP and OAMP by varying the condition number in application of compressed sensing. As can be observed from this figure, MAMP and OAMP converge to the same fixed point. In $\kappa(\bs{H})=1$, MAMP has the comparable convergence speed as OAMP. However, as the $\kappa(\bs{H})$ increases, MAMP need to pay more iteration times to converge the same fixed point as OAMP. Also, we note that the convergence speed and NMSE performance of MAMP and OAMP tend to worse in large condition number.

\section{Conclusions}
In this paper, we reviewed several AMP-like algorithms: AMP, OAMP, VAMP, and MAMP. We began at introducing AMP algorithm, which is originally proposed for providing a sparse solution to LASSO inference problem but  is widely  applied to a lot of engineering fields under Bayes-optimal setting. In IID sub-Gaussian random measurement matrices region, the AMP algorithm can achieve Bayes-optimal MSE performance, but it may fail to converge if random measurement is beyond IID sub-Gaussian. Following AMP, we introduced a modified AMP algorithm termed OAMP, which modified AMP in two aspects:  LMMSE de-correlated matrix and divergence-free denoiser. The OAMP algorithm can be applied to more general region: unitarily-invariant matrix, but it should be payed more computational complexity due to matrix inversion. To balance the computational complexity and random measurement region, the MAMP algorithm applies several terms of matrix Taylor series to approximate matrix inversion and applies all preceding messages to ensure three orthogonality. The MAMP algorithm relies on the given spectral of sample of random measurement matrix. Although, several works gave some approximations to it, the complexity is still huge. In addition, the convergence speed of MAMP is slower than OAMP especially in the case of large condition number. On the other hand,  a significant feature of AMP-like algorithms is that their asymptotic MSE performance can be fully predicted by their SEs. We also gave a brief derivation of their SEs.

\section{Acknowledgements}
We are  grateful to Y. Kabashima, D. Cai, and Y. Fu for valuable comments and useful discussions.

\bibliographystyle{IEEEtran}
\bibliography{ZQY_bib}

\end{document}